\def\version{August 25, 2005}

\documentclass[12pt]{article}
\usepackage{latexsym}
\usepackage{amsmath}
\usepackage{amsfonts}
\usepackage{amssymb}
\usepackage{graphicx}

\numberwithin{equation}{section}

\begin{document}

\begin{titlepage}
\title{String Matching and 1d Lattice Gases}

\author{Muhittin Mungan \\
Bo\u gazi\c ci University, \\
Department of Physics, \\
34342 Bebek, Istanbul, Turkey} 
\date{\version}
\maketitle \thispagestyle{empty}

\begin{abstract}
{\footnotesize We calculate the probability distributions for the number of 
occurrences $n$ of of a given $l$ letter word in a random string of $k$ letters. 
We consider the case in which the letters of the strings belong to 
an $r$ letter alphabet. Analytical expressions for the distribution are 
known for the asymptotic regimes (i) $k \gg r^l \gg 1$ (Gaussian) and 
$k,l \rightarrow \infty$ such that $k/r^l$ is finite (Compound Poisson). 
However, it is known that these distributions do now work well in the 
intermediate regime $k \gtrsim r^l \gtrsim 1$. 
We show that the problem of calculating the string matching probability 
can be cast into a problem of determining the 
configurational partition function of a 1d lattice 
gas with interacting particles such that the string matching 
probability distribution becomes the grand-partition sum 
of the lattice gas, with the number of particles corresponding 
to the number of matches on the string. Using this analogy, we perform a 
virial expansion of the effective equation of state and 
thereby obtain the probability distribution function. 
Our result reproduces the behavior of the matching distribution in all regimes, 
{\it i.e.} the asymptotic as well as the intermediate regimes, rather well. 
We are also able to show analytically how the limiting distributions arise. 
Our analysis builds on the observation that the effective interactions between 
the particles consist of a relatively strong core of size $l$, the word length, 
followed by a weak, exponentially decaying tail, 
whose overall strength decreases with increasing $l$. 
We find that the asymptotic regimes correspond to the case where the 
tail of the interactions can be neglected, while in the intermediate regime 
the effects of the tail needs to be incorporated into the analysis. 
This is ultimately responsible for the failure of the asymptotic 
distributions in this regime. Our results are readily generalized to the 
case where the random strings are generated by more complicated stochastic 
process such as a non-uniform letter probability distribution or Markov chains.
We show that by varying the parameters of the stochastic process, the tails of the 
effective interactions can be made even more dominant rendering thus the 
asymptotic approximations less accurate in such a regime.

}
\end{abstract}
\end{titlepage}

\section{Introduction}

The problem of determining the probability of encountering (matching) 
a given string of length $l$ in another string of length $k$, 
whose letters have been drawn randomly from 
an alphabet of $r$ letters, has a variety of applications 
ranging from designing fast algorithms for pattern searching \cite{boyer,knuth},
to problems in genetics such as assessing the likelihood of events
such as the frequency of 
occurrence of DNA segments \cite{pevzner,prum}, or that certain  DNA segments 
align \cite{karlin,dembo}. In each of theses cases, the likelihood 
estimates for random sequences can be used as a benchmark against which one
can evaluate the statistical significance of actually observed events.

The problem is non-trivial, because of the possibility of overlapping
occurrences in the string, which introduce correlations that need to be
dealt with. Guibas and Odlyzko \cite{guibas0,guibasI,guibasII} 
derived the moment generating functions associated with 
the probability for not encountering a given set of words in a random string, 
whose letters were distributed independently and identically. The resulting
distributions turn out to depend on a set of correlation functions that 
capture the overlap properties of the words with each other. 

Building on the work of Guibas and Odlyzko, several authors have studied the 
probability distribution for the number of occurrences $n$ of of a given 
$l$ letter word in a random string of $k$ letters, under various 
assumptions on the distribution of random letters 
\cite{chrys,geske,fudos,regszpan,pevzner,goldwater,waterman,schbath,prum,reinertetal}: 
The  cases where the letters of the random string are independently 
and identically distributed (i.i.d.) was treated by Fudos et al. \cite{fudos}, 
whereas the case where the letter distribution follows the 
steady state distribution of a Markov process has been 
investigated by several authors \cite{chrys,geske,regszpan,waterman,reinertetal}.
All of these results have been obtained for asymptotic regimes 
($k$ large + various assumptions on the length of the word $l$), 
where tools of statistics such as the central limit theorem 
\cite{fudos,regszpan,prum}, theory of large deviations 
\cite{reinertetal,regszpan}, or (compound) poisson 
approximations for rare-events \cite{chrys,geske,goldwater,schbath} are 
applicable. 

The regimes of applicability can be difficult to identify, however. 
It has been noted that, even in the case of i.i.d letters, when the 
length $l$ of the word to be matched is fixed, and assuming the length 
of the random string to be large, the most accurate approximation 
to chose (gaussian or compound poisson) still depends on the 
word itself that is being matched \cite{robschbath}.  
  
It is therefore desirable to come up with a single, explicit 
analytical expression for the 
probability distribution  that is generally valid, and to obtain the 
asymptotic expressions, mentioned above, as special cases by 
taking certain limits. 
This is what we set out to do in this article. Besides the 
obvious advantage of having a single description, such an approach will 
naturally identify the regimes of application of the various asymptotic
approximations, while also pointing out when and how they fail.  

It turns out that all of these issues are present even in the simplest 
case where the letters of the random string are uniformly and independently
distributed.  For the sake of simplicity and clarity of presentation, we 
will perform the analysis for this case. However, we will point out 
in detail how these results carry over to the more general case of 
random letter distributions.  
  
Our approach to this problem, which appears to be novel, 
can be summarized as follows: We first show that the 
problem of calculating the probability distribution for the number of 
occurrences $n$ of of a given $l$ letter word in a random string of $k$ 
letters, can be rigorously mapped into the problem of calculating the 
configurational part of the grand-canonical partition function 
of a 1d lattice gas. In this
mapping  the number of particles correspond to the number occurrences,
the "volume" of the gas is the length of the random string, and the correlations 
between subsequent occurrences turn into pairwise interactions whose nature 
depends on properties of the word to be matched. It turns out that common to
all interactions is a relatively strong and short-ranged segment of range $l$, 
followed by a weak and exponentially decaying tail. 
 
With the help of the lattice gas analogy, and by using techniques of 
liquid theory, such as the virial expansion, we are able to obtain 
an analytical expression for the probability distribution that 
reproduces the known asymptotic limits. We show how the distribution
crosses over into the asymptotic forms of the distribution, 
and thereby expose the conditions required 
for these limits to be applicable.  

More importantly, our method allows us to analytically treat the 
intermediate regime of moderate string lengths, $k \gtrsim r^\ell \gtrsim 1$, 
as well. This regime is most relevant for biological applications and 
turns out to be the hardest one to tackle analytically,
since in this regime the effects of the tail are strong and need to be kept 
in the analysis. This is also the  reason behind the deviations of the 
asymptotic forms from the actual distribution, since, as we will show, 
these distributions are obtained by neglecting the tail of the interactions. 
These deviations become more pronounced for short words and
small number of letters in the alphabet, small $l$ and $r$, respectively. 

Our results are readily generalized to the broader class of letter distributions,
such as non-uniformly distributed letters or letters generated by a Markov 
process. Such distributions give rise to a broader class of 
effective interactions. In particular, it turns out that these 
interactions can have  stronger tails than can be achieved by a 
uniform letter distribution. This potentially renders our method 
of approach even more relevant to such letter distributions. 
  
We would also like to note that our approach is in spirit similar to recent 
attempts at solving combinatorial problems, such as the k-SAT problem 
\cite{kirkpatrick,mezard,mertens,achlioptas}, using ideas borrowed from 
statistical mechanics \cite{fuanderson,monasson}.
     
The article is organized as follows: 
In Section II we introduce our notation and formalism, 
rederiving in this setting some of the relevant and known results. 
In Section III we establish the partition function analogy. 
We derive and study the properties of the effective particle
interactions and then set up a virial expansion for the 
"equation of state". From the virial expansion we obtain 
the $n$-particle partition function, which in this analogy 
corresponds to the $n$-match probability distribution function. 
We show how the various know limits
arise and discuss the underlying assumptions. 
Section IV discusses the generalization and implications of our 
approach to the more general class of letter distributions studied
in the literature and we will discuss our 
results in Section V.  
    
\section{The matching probabilities}
In this section we derive some of the known expressions for the 
the matching and $n$-match probability, the probability of a
at least one and precisely $n$ occurrences of a given word, respectively.
Besides providing a review of the relevant results, the main 
purpose of this section is to introduce our notation and 
provide the setting for the statistical mechanics approach
to be taken up in the following Section.  
   
\subsection{Definitions}

Assume that $x$, and $y$ are variables that take
values from an $r$ letter alphabet such that
$x, y \in \{0, \ldots , r-1\} $. 
Let ${\bf x} = (x_1,x_2,x_3, \ldots, x_{l})$ and 
${\bf y} = (y_1,y_2,y_3, \ldots, y_{l})$,
be two strings of $l$ letters. 
Define the match indicator function  
$ \Phi({\bf x},{\bf y})$ as
\begin{equation}
\Phi({\bf x},{\bf y }) = \prod_{t=1}^{l} \delta(x_t,y_t) 
\label{eqn:Phidef}
\end{equation}
So that we have
\begin{equation}
\Phi({\bf x}, {\bf y}) =
\left \{\begin{array}{ll}  1, & \mbox{if ${\bf x} = {\bf y}$} \\
                                      0,   & \mbox{otherwise}. \end{array} \right. 
\label{eqn:Phidefetaz}
\end{equation}

Let ${\bf y} = (y_1,y_2, \ldots , y_k) $ be a string of length 
$k \ge l$ and denote by ${\bf y}_{a,l} = ( y_{a+1}, y_{a+2}, \ldots , y_{a+l})$
the substring of length $l$ starting at position $a$, $ a = 0, 1, \ldots, k-l$. 
Furthermore, let
\begin{equation}
f_a({\bf x},{\bf y}) = \Phi({\bf x},{\bf y}_{a,l}).   
\label{eqn:fdef}
\end{equation}
We have
\begin{equation}
f_a({\bf x},{\bf y}) 
 = \left \{\begin{array}{ll}  1, & \mbox{$ {\bf x} = {\bf y}_{a,l}$ } 
          \\ 0,   & \mbox{otherwise}. \end{array} \right.
\label{eqn:fadef}
\end{equation}
In other words, $f_a({\bf x},{\bf y}) = 1$,
if and only if ${\bf x}$ matches ${\bf y}$ at position $a$, and zero otherwise.

\subsection{The matching probability}

Define $p(m; { \bf x})$ to be the the probability  that 
a given word ${\bf x}$ of length $l$ is contained 
{\em at least once} in a randomly 
drawn string ${ \bf y}$ of length $m+l$. We will refer to this as the 
{\em matching probability}.
Let $I_M({\bf x},{\bf y})$ be the function that takes on the value
one if the $k$-string ${ \bf y}$ contains the given $l$-string ${\bf x}$ 
at least once, and
zero otherwise. Using Eq.~(\ref{eqn:fadef}), we can write
\begin{equation}
I_M({\bf x},{\bf y}) = 1 -  \prod_{a=0}^{k-l}
\left [ 1 - f_a({\bf x},{\bf y}) \right ].
\end{equation}
Since $k \ge l$, it is convenient to define the excess length $m = k-l$. 
We thus find
\begin{equation}
p(m; { \bf x}) = 1 - \frac{1}{r^{m+l}} \sum_{{\bf y}}^{} \prod_{a=0}^{m}
\left [ 1 - f_a({\bf x},{\bf y}) \right ],
\label{eqn:p1function}
\end{equation}
where $r^{m+l}$ is the number of distinct $k=m+l$-strings of $r$-letters, 
and the summation is over all such strings ${\bf y}$.

In \cite{mungan}, 
the products on the right hand side of Eq.~(\ref{eqn:p1function}) 
were expanded into a Mayer-like sum, 
\begin{equation}
p(m;{\bf x}) = \frac{1}{r^k} \sum_{{\bf y}}^{} \sum_{a} f_a  
 - \frac{1}{r^k} \sum_{{\bf y}}^{} \sum_{a < b} f_a f_b 
  + \frac{1}{r^k} \sum_{{\bf y}}^{} \sum_{a < b < c} f_a f_b f_c - \ldots,
\label{eqn:meier}
\end{equation}
(arguments of $f_a$ will be suppressed in what follows)
and the terms in the sum where evaluated approximately. 
Here we will take a different approach. 

The following algebraic identity will be of use in the following:
\begin{equation}
1 - \prod_{a=0}^{m} \left ( 1 - f_a \right ) = 
\sum_{b=0}^{m} f_b  \prod_{a=0}^{b-1} \left ( 1 - f_a \right ),  
\label{eqn:flemma}
\end{equation}
with the convention that when $b=0$, 
the product on the right hand side is set to one. 
Eq.~(\ref{eqn:flemma}) is readily proven by induction.

Using this identity, $p(m;{\bf x})$ can be written as
\begin{equation}
p(m;{\bf x}) = \frac{1}{r^{m+l}} \sum_{{\bf y}}^{} 
\sum_{b=0}^{m} f_b  \prod_{a=0}^{b-1} \left ( 1 - f_a \right ).
\end{equation}
Note that for any
given $b$, the expression on the right hand side only involves the variables
$y_1, y_2, \ldots, y_{b+l}$. The sum over the remaining indices yields 
$r^{m-b}$ and we find that
\begin{equation}
p(m;{\bf x}) = \sum_{b=0}^{m} \frac{1}{r^{b+l}} 
\sum_{y_1 \cdots y_{b+l}}^{} f_b  \prod_{a=0}^{b-1} \left ( 1 - f_a \right ).
\end{equation} 
Defining the correlator $d(b;{\bf x})$ as
\begin{equation}
d(b;{\bf x}) = \sum_{y_1 \cdots y_{b+l}}^{} f_b  \prod_{a=0}^{b-1} \left ( 1 - f_a 
\right ),
\label{eqn:ddef}
\end{equation}
$p(m;{\bf x})$ can be therefore written as
\begin{equation}
p(m;{\bf x}) = \sum_{b=0}^{m} \frac{1}{r^{b+l}} d(b;{\bf x}).
\label{eqn:pdef}
\end{equation}

We can obtain a recursion relation for $d(b;{\bf x})$ by factoring 
out the $a=0$ term in Eq.~(\ref{eqn:ddef}),  
\begin{equation}
d(b;{\bf x}) = \sum_{y_1 \cdots y_{b+l}}^{} f_b  \prod_{a=1}^{b-1} 
\left ( 1 - f_a \right ) -
\sum_{y_1 \cdots y_{b+l}}^{} f_0 f_b  \prod_{a=1}^{b-1} \left ( 1 - f_a \right ).
\end{equation}
The argument of the first sum does not contain the variable $y_1$,
while the sum over the remaining variables yields $d(b-1;{\bf x})$. Thus,
\begin{equation}
d(b;{\bf x}) = r d(b-1;{\bf x}) - h(b;{\bf x}),
\label{eqn:drec}
\end{equation}
with the correlator $h$ defined as  
\begin{equation}
h(b;{\bf x}) = \sum_{y_1 \cdots y_{b+l}}^{} f_0 \left [ \prod_{a=1}^{b-1} 
\left ( 1 - f_a \right ) \right ] f_b.
\label{eqn:gdef}
\end{equation}

Note that for $m=0$
\begin{equation}
p(0; {\bf x}) = \frac{1}{r^l}.
\end{equation}
Comparing with Eq.~(\ref{eqn:pdef}) this implies that
\begin{equation}
d(0; {\bf x}) = 1.
\end{equation}
Since there are no constraints on $h(0; {\bf x})$, we will define
$h(0; {\bf x}) = 0$.

We next seek a recursion relation for $h$. Using the
identity, Eq.~(\ref{eqn:flemma}), we find from Eq.~(\ref{eqn:gdef})
\begin{equation}
h(b;{\bf x}) =  
\sum_{y_1 \cdots y_{b+l}}^{} \left \{ f_0 f_b - \sum_{c=1}^{b-1} f_0 
\left [ \prod_{a=1}^{c-1} \left ( 1 - f_a \right ) \right ] f_c f_b  \right \}.
\label{eqn:g2}
\end{equation}
Recall from the definition of $f_b({\bf x},{\bf y})$ that $f_b$ is a product of 
Kronecker deltas, Eqs.~({\ref{eqn:Phidefetaz}) and
({\ref{eqn:fdef}). The Kronecker deltas enforce a transitive relation between their 
 arguments, and we can write 
$f_b({\bf x},{\bf y};0) = f_b({\bf x},{\bf y};0) f_b({\bf x},{\bf \tilde{y}};0)$,
where we have introduced an auxiliary set of variables ${\bf \tilde{y}}$ over 
which a sum is to be performed. 
Thus Eq.~(\ref{eqn:g2}) can be rewritten as
\begin{equation}
\sum_{y_1 \cdots y_{b+l}}^{} \sum_{c=1}^{b-1} f_0 \left [ \prod_{a=1}^{c-1} 
\left ( 1 - f_a \right ) \right ] f_c f_b = 
\sum_{c=1}^{b-1} \left \{ \sum_{y_1 \cdots y_{b+l}}^{}  f_0 \left [ \prod_{a=1}^{c-1} 
\left ( 1 - f_a \right ) \right ] f_c \right \}  
 \left \{ \sum_{\tilde{y}c \cdots \tilde{y}_{b+l}}^{} 
f_c f_b \right \}.
\label{eqn:grel}
\end{equation}
Defining the correlator $C(b;{\bf x})$ as
\begin{equation}
C(b;{\bf x}) = \sum_{y_1 \cdots y_{b+l}}^{} f_0({\bf x},{\bf y}) 
f_b({\bf x},{\bf y}),
\label{eqn:cdef} 
\end{equation}
and substituting Eq.~(\ref{eqn:grel}) into 
Eq.~(\ref{eqn:g2}), we find 
\begin{equation}
h(b;{\bf x}) = C(b;{\bf x}) - \sum_{a=1}^{b-1} h(a;{\bf x}) C(b-a;{\bf x}).	
\label{eqn:grec} 
\end{equation}
Using Eq.~(\ref{eqn:cdef}), it can be easily shown that for $b\ge l$
\begin{equation}
C(b;{\bf x}) = r^{b-l}.
\label{eqn:CCdef}
\end{equation}

Denoting the values of $C(b;{\bf x})$ for $b<l$ by $c_b({\bf x})$, we have 
\begin{equation} 
c_b({\bf x}) = \sum_{y_1 \cdots y_{b+l}}^{} 
f_0({\bf x},{\bf y};0) f_b({\bf x},{\bf y};0), \; \; 0 < b < l, 
\label{eqn:cexpl} 
\end{equation}
and thus 
\begin{equation}
C_b = \left \{ \begin{array}{ll} c_b, & \mbox {$0 < b < l$,} \\ r^{b-l} & \mbox{$b \ge l$.} 
\end{array} \right.
\label{eqn:Concemore}
\end{equation}

As evident from Eq.~(\ref{eqn:cexpl}), the set of indices $c_b({\bf x}) \in \{0,1 \}$, 
with $b=1,2,\ldots,l-1$ measure the auto-correlations of ${\bf x}$. They 
are referred to as the bit-vector ${\bf c} = (c_1,c_2,\ldots,c_{l-1})$ 
associated with ${\bf x}$, and were studied by Harborth \cite{harborth} and 
later in considerable detail by Guibas and 
Odlyzko \cite{guibas0,guibasI,guibasII}. 

\subsection{Bit-vectors}
\label{bit-sect}
From the definition, Eq.~(\ref{eqn:cexpl}), it is 
clear that $c_b = 1$ if and only if the string ${\bf x}$ shifted by an amount 
$b$ relative to itself coincides on the overlapping part. Conversely $c_b = 0$,
if the overlapping part does not coincide. It turns out that the set
of $r^l$ possible words ${\bf x}$ of length $l$ are partitioned into equivalence
classes with respect to their bit-vectors ${\bf c} = (c_1,c_2, \ldots c_{l-1})$ 
and that the possible classes are independent of the number of letters $r$ 
(as long as $r \ge 2$) \cite{guibasI}.
Tables \ref{bitv1} and \ref{bitv2} list the sets of possible 
bit-vectors upto $l=8$ along with the number of elements in their respective 
equivalence classes for $r=2,3,4$. 

We see that the definition of $c_b({\bf x})$ imposes strong 
conditions on the possible values of the $l-1$ bits of a bit-vector and it turns out 
that the resulting bit-vectors have interesting properties \cite{guibasI,rivrah},
of which we will mention only the most relevant ones.

For example, if $c_p = c_q = 1$ with $p<q$ this implies that $c_t = 1 $ for all $t$ of the 
form $t = p + i (q-p)$ with, $i=0, 1, 2, \ldots$ and  $t < l$. This is referred
to as the {\em forward propagation rule} \cite{guibasI}. In particular, 
$c_p = 1$ implies that $c_{ip} = 1$ for all $i, 1, 2,  \ldots$ such that 
$ip < l$. The latter result shows that $p$ can be considered as a period.
We define the {\em fundamental period} of a string ${\bf x}$, $\chi({\bf x})$,
to be the smallest $p$, with $0<p<l$ such that $c_p = 1$. If $ {\bf x}$ is such
that its bit-vector is $000\cdots0$ (all zeroes), we define $\chi({\bf x}) = l$.     

\subsection{The $n$-match probability}

Denote  by $p(n;m,{ \bf x})$ the probability that 
that a randomly drawn $k$-string ${ \bf y}$ contains a given $l$-string 
${\bf x}$ {\em precisely} $n$ times. We will refer to this as the 
{\em n-match probability}.

If we let the random variable $N({\bf x},{\bf y})$ denote the number 
of occurrences of ${\bf x}$ in ${\bf y}$, it follows that 
\begin{equation}
N({\bf x},{\bf y}) = \sum_{a=0}^{m} f_a.
\end{equation}
Thus the average number of matches $\langle n \rangle$ and its 
second moment $\langle n^2 \rangle$ are readily obtained as
\begin{equation}
\langle n \rangle = \frac{1}{r^{m+l}} \sum_{\bf y} \sum_{a=0}^{m} f_a = \frac{m+1}{r^l} 
\label{eqn:nave}
\end{equation}
and
\begin{equation}
\langle n^2 \rangle = \frac{1}{r^{m+l}} \sum_{\bf y} \sum_{a<b}^{} f_a f_b.
\end{equation}

\twocolumn
\begin{table}
\vskip 0.25cm
\begin{tabular}{||r|l|l|l||}
\hline
${\bf c}$  & $r=2$    & $r=3$ & $r=4$  \\ 
\hline
0 & 2 & 6 & 12 \\ 
1 & 2 & 3 & 4 \\ 
\hline
00 & 4 & 18 & 48 \\ 
01 & 2 & 6 & 12 \\ 
11 & 2 & 3 & 4 \\ 
\hline
000 & 6 & 48 & 180 \\ 
001 & 6 & 24 & 60 \\ 
010 & 2 & 6 & 12 \\ 
111 & 2 & 3 & 4 \\ 
\hline
0000 & 12 & 144 & 720 \\ 
0001 & 10 & 66 & 228 \\ 
0010 & 4 & 18 & 48 \\ 
0011 & 2 & 6 & 12 \\ 
0101 & 2 & 6 & 12 \\ 
1111 & 2 & 3 & 4 \\ 
\hline
00000 & 20 & 414 & 2832 \\ 
00001 & 22 & 210 & 948 \\ 
00010 & 6 & 48 & 180 \\ 
00011 & 6 & 24 & 60 \\ 
00100 & 4 & 18 & 48 \\ 
00101 & 2 & 6 & 12 \\ 
01010 & 2 & 6 & 12 \\ 
11111 & 2 & 3 & 4 \\ 
\hline
\end{tabular}
\caption{Equivalence classes of bit-vectors and their number of elements. 
The table shows the bit-vectors ${\bf c} = (c_1,c_2,\ldots,c_{l-1})$
associated with strings of length $l=2,3,4,5$ and $6$ and 
the number of elements in these equivalence classes 
for $r=2,3$ and $4$ letter alphabets.}    
\label{bitv1}
\end{table}

\begin{table}
\vskip 0.25cm
\begin{tabular}{||r|l|l|l||}
\hline
${\bf c}$  & $r=2$    & $r=3$ & $r=4$  \\ 
\hline
000000 & 40 & 1242 & 11328 \\ 
000001 & 38 & 606 & 3732 \\ 
000010 & 16 & 162 & 768 \\ 
000011 & 12 & 72 & 240 \\ 
000100 & 8 & 54 & 192 \\ 
000101 & 2 & 12 & 36 \\ 
000111 & 2 & 6 & 12 \\ 
001001 & 6 & 24 & 60 \\ 
010101 & 2 & 6 & 12 \\ 
111111 & 2 & 3 & 4 \\ 
\hline
0000000 & 74 & 3678 & 45132 \\ 
0000001 & 82 & 1866 & 15108 \\ 
0000010 & 26 & 462 & 3012 \\ 
0000011 & 22 & 210 & 948 \\ 
0000100 & 16 & 162 & 768 \\ 
0000101 & 8 & 54 & 192 \\ 
0000111 & 6 & 24 & 60 \\ 
0001000 & 6 & 48 & 180 \\ 
0001001 & 6 & 24 & 60 \\ 
0010010 & 4 & 18 & 48 \\ 
0010011 & 2 & 6 & 12 \\ 
0101010 & 2 & 6 & 12 \\ 
1111111 & 2 & 3 & 4 \\ 
\hline
\end{tabular}
\caption{Equivalence classes of bit-vectors and their number of elements. 
The table shows the bit-vectors ${\bf c} = (c_1,c_2,\ldots,c_{l-1})$
associated with strings of length $l=7$ and $8$ and 
the number of elements in these equivalence classes 
for $r=2,3$ and $4$ letter alphabets.}    
\label{bitv2}
\end{table}

\onecolumn

The latter expression can be worked out using Eqs.~(\ref{eqn:cdef}), (\ref{eqn:CCdef}) and 
(\ref{eqn:cexpl}) and we find for the variance
\begin{equation}
\sigma_n^2 = \frac{m+1}{r^l} + \frac{1}{r^{2l}} \left [ (m+1)(1-2l) + l(l-1) \right ] 
+ \frac{1}{r^l} \sum_{b=1}^{l-1} (m+1-b)c_b({\bf x}).
\label{eqn:sigman}
\end{equation}
Eq.~(\ref{eqn:sigman}) is a special case of a result due to Kleffe and Borodovsky 
\cite{kleffe}, who considered general distributions of random letters. 

Let 
$I_{n,m}(a_1,a_2,\ldots,a_n;{\bf x}, {\bf y})$ be the
function that takes on the value $1$ when $n$ matches occur that are
located at positions
$a_1,a_2,\ldots,a_n$, with $0 < a_1 <a_2  < \cdots < a_n < m$ and zero otherwise,
\begin{eqnarray}
I_{n,m}(a_1,a_2,\ldots a_n;{\bf x}, {\bf y}) &=& \nonumber \\
\left [ \prod_{i_1=1}^{a_1-1}(1-f_{i_1}) \right ] f_{a_1} 
 \left [ \prod_{i_2=a_1+1}^{a_2-1}(1-f_{i_2}) \right ] f_{a_2} 
 &\cdots&
 \left [ \prod_{i_n=a_{n-1}+1}^{a_n-1}(1-f_{i_n}) \right ] f_{a_n} 
 \left [ \prod_{i_{n+1}=a_{n}+1}^{m}(1-f_{i_{n+1}}) \right ] . \nonumber \\
\label{eqn:Imndef}  
\end{eqnarray}
In terms of  $I_{n,m}(a_1,a_2,\ldots a_n;{\bf x}, {\bf y})$ we can 
write $p(n;m,{\bf x})$ as
\begin{equation}
p(n;m,{\bf x}) = 
\sum_{a_1<a_2<\cdots<a_n}\frac{1}{r^{m+l}}\sum_{\bf y} 
I_{n,m}(a_1,a_2,\ldots a_n;{\bf x}, {\bf y}). 
\label{eqn:pnmdef}
\end{equation}
 
Analogously to the reasoning leading from Eq.~(\ref{eqn:g2}) to 
Eq.~(\ref{eqn:grel}), it can be shown that the sum over ${\bf y}$ factorizes 
$I_{n,m}(a_1,a_2,\ldots a_n;{\bf x}, {\bf y})$ as 
\begin{equation}
\frac{1}{r^{m+l}} \sum_{\bf y} I_{n,m}(a_1,a_2,\ldots a_n;{\bf x}, {\bf y}) 
= \frac{1}{r^{m+l}} d(a_1) 
\left [ \prod_{i=1}^{n-1}h(a_{i+1}-a_{i}) \right ] d(m-a_n),
\label{eqn:Gaaexp}
\end{equation}
where $d$ and $h$ are as defined in Eqs.~(\ref{eqn:ddef}) and (\ref{eqn:gdef}). Thus 
$p(n;m,{\bf x})$ becomes
\begin{equation}
p(n;m,{\bf c})
= \sum_{a_1<a_2<\cdots<a_n} \frac{1}{r^{m+l}} d(a_1) 
\left [ \prod_{i=1}^{n-1}h(a_{i+1}-a_{i}) \right ] d(m-a_n),
\label{eqn:pnmaspartition}
\end{equation}
where we have changed the argument of the distribution function to 
$p(n;m,{\bf c})$, to emphasize that the
distribution really depends on the bit-vector ${\bf c}$ only. It is readily seen 
that the sum over the positions $a_i$ is an $n+1$ fold convolution of $d$ and $h$.
To simplify the results as well as well as to be able to obtain 
asymptotic expressions, we next introduce generating functions.

\subsection{Generating functions}

Define the generating function $g(z)$ associated with a sequence $g(b)$ by
\begin{equation}
g(z) = \sum_{b=0}^{\infty} z^b g(b).
\end{equation}

From Eqs.~(\ref{eqn:cdef}), (\ref{eqn:CCdef}) and (\ref{eqn:cexpl}) we find
\begin{equation}
C(z;{\bf x}) = c(z;{\bf x}) + \frac{z^l}{1-zr},
\label{eqn:czdef} 
\end{equation}
where $c(z;{\bf x})$ is a polynomial of degree $l-1$, 
\begin{equation}
 c(z;{\bf x}) = \sum_{b=1}^{l-1} z^b c_b({\bf x}). 
\end{equation}
It is useful to also define the polynomial of degree $l$, $\lambda(z;{\bf c})$, as 
\begin{equation}
\lambda(z;{\bf c}) = z^l + r^l(1-z) \left [ 1 + c(z/r;{\bf x}) \right ] .
\label{eqn:lambdazdef}
\end{equation}

Using Eqs.~(\ref{eqn:drec}),(\ref{eqn:grec}), (\ref{eqn:CCdef}) and (\ref{eqn:lambdazdef}), 
we see that the generating function of $h$ and $d$ are given in terms of $c(z/r;{\bf x})$ as
\begin{equation}
h(z/r;{\bf c}) = 1 - \frac{1}{1 + C(z/r;{\bf x})} = 1 - \frac{1}{1+ c(z;{\bf x}) +\frac{1}{r^l} \frac{z^l}{1-z}},
\label{eqn:hexpl0}
\end{equation}
which can be written in terms of $\lambda(z;{\bf c})$ as
\begin{equation}
h(z/r;{\bf c}) = 1 - r^l \; \frac{1 - z}{\lambda(z;{\bf c})},
\label{eqn:hzexplicit} 
\end{equation}
and likewise, 
\begin{equation}
d(z/r;{\bf c}) = \frac{1 - h(z/r;{\bf c})}{1-z} = \frac{r^l}{\lambda(z;{\bf c})}.
\label{eqn:dzexplicit} 
\end{equation}

The generating function for $p(m;{\bf c})$ thus becomes
 \cite{guibas0,guibasII}
\begin{equation}
p(z;{\bf x}, 0) = \frac{1}{1-z} \; \; 
\frac{1}{\lambda(z;{\bf c})}. 
\label{eqn:pzex}
\end{equation}

Turning next to the generating function of
$p(n;m,{\bf c})$, 
\begin{equation}
p(n;z,{\bf c}) = \sum_{m=n}^{\infty} z^m p(n;m,{\bf c}),
\end{equation}
we obtain (for $n \ge 1$)
\begin{equation}
p(n;z,{\bf c}) = \frac{1}{r^l} d(z/r;{\bf c})^2 h(z/r;{\bf c})^{n-1}.
\label{eqn:pnzdef}
\end{equation}
Eq.~(\ref{eqn:pnzdef}) is a special case of a more general result due to R\'egnier 
and Szpankowski \cite{regszpan} who consider a broader class of 
letter distributions,  
including inhomogeneous letter distributions as well as sequences of 
random letters generated from the steady-state of a Markov process. 

As an aside, we can alternatively write \cite{regszpan} $p(n;z,{\bf c})$ 
in terms $c(z/r)$ alone as 
\begin{equation}
p(n;z,{\bf c}) = \frac{1}{r^l} \; \; \frac{\left [ r^l(1-z)c(z/r) + z^l \right ]^{n-1}}
{\left [ r^l(1-z)(1+c(z/r)) + z^l \right ]^{n+1}},
\end{equation}
or in terms of the matching probability $p(z;{\bf c})$ as
\begin{equation}
p(n;z,{\bf c}) =  r^l (1-z)^2p(z;{\bf c})^2 
\left [ 1 - r^l (1-z)^2 p(z;{\bf c}) \right ]^{n-1}.
\end{equation}
Note that from the last expression, we recover again Eq.~(\ref{eqn:pzex})
in terms of the generating functions, 
\begin{equation}
\sum_{n=1}^{\infty} p(n;z,{\bf c}) = p(z;{\bf c}). 
\end{equation}
In fact for $n=0$ we therefore have
\begin{equation}
p(0;z,{\bf c}) = \frac{1}{1-z} - p(m;{\bf c}) 
=  \frac{1}{1-z} \; \left [ 1 - \frac{1}{\lambda(z;{\bf c})} \right ].
\end{equation}
 
\subsection{Asymptotic behavior}
\label{asySec}
Once the generating functions have been determined, the original functions can be obtained
by an inverse transformation defined as follows:
If $f(z)$ is the generating function associated with $f(b)$, then
\begin{equation}
f(b) = \frac{1}{2\pi i} \oint_{\partial D} dz \frac{1}{z^{m+1}} \; \; f(z),
\label{eqn:fbycontour}
\end{equation}
where $\partial D$ is the boundary of a domain $D$ in the complex plane that includes 
the origin and on which $f(z)$ is analytic \cite{wilf}.

Note that the generating functions of 
$h(z;{\bf c})$, $d(z;{\bf c})$, $p(z;{\bf c})$, and $p(n;z,{\bf c})$ 
are all rational functions, with their denominators involving $\lambda(z;{\bf c})$ 
or its powers and that they all go to zero as $\| z \| \rightarrow \infty$. 
For example, for the matching probability we have
\begin{equation}
p(m,{\bf c}) = \frac{1}{2\pi i} \oint_{\partial D} dz \frac{1}{z^{m+1}} \; \;
 \frac{1}{(1-z)\lambda(z;{\bf c})}.
\label{eqn:pmzsol}
\end{equation} 
As we will show below, the behavior of $p(m,{\bf c})$ for large $m$ (and likewise for
$h$, $d$ and $p(n;m,{\bf c}$) is dominated by the zeroes of $\lambda(z;{\bf c})$ 
that are closest to the origin.

A numerical inspection of the zeroes of $\lambda(z;{\bf c})$
for $2< l < 10$ and $r=2,3,4$ shows that: (1) All zeroes $z_i$  of $\lambda$ are distinct, 
(2) the zero of smallest magnitude, $z_1$, is real, and greater but 
near $1$ and (3) all other zeroes have
magnitudes of the order $\| z_i\| \sim r, i = 2, \ldots l $. Fig. \ref{rootplot}
shows a plot of the zeroes for $l=4,r=2$ and $l=8,r=2,3,4$. In fact, it can be 
rigorously proven \cite{guibas0} that $\lambda(z;{\bf c})$ has a single (real) 
zero in a circular domain centered at $z=1$ and of sufficiently small 
radius $\epsilon$. 

The asymptotic behavior of $f(b)$ in Eq.~(\ref{eqn:fbycontour}) can be 
obtained by stretching the contour $\partial D$ to infinity while 
circling around the zeroes of $f(z)$ without including them. The integral
over the boundary at infinity turns out to yields no contribution since for the 
cases of interest $f(z) \rightarrow 0$ and the integrand is 
asymptotically of the order of at least $1/z^{m+1}$. 
This leaves the contributions from the zeroes of $(1-z)\lambda(z;{\bf c}) $, which 
are traversed counter-clockwise, if the contour at infinity is traversed clock-wise. 

\begin{figure}[!ht]
\begin{center}
\end{center}
\includegraphics[width=16cm]{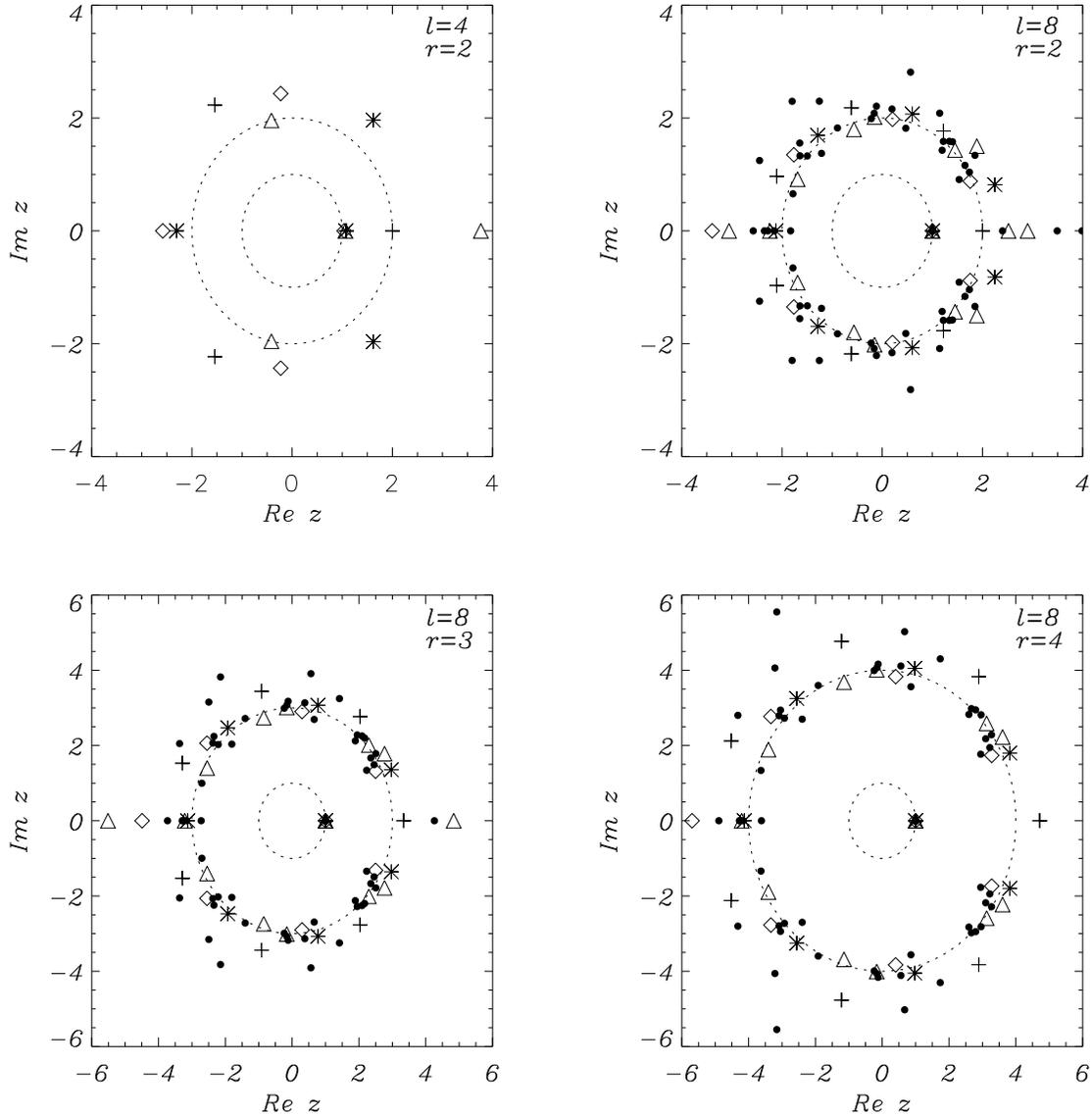}
\caption[]{Root Loci of the polynomial $\lambda(z;{\bf c})$, Eq. (\ref{eqn:lambdazdef}). 
The figures are for $(l,r)$ - values (starting from the top left and going 
clockwise) $(4,2)$, $(8,2)$, $(8,3)$ and $(8,4)$ . 
Plotted in each figure are the roots associated with the possible equivalence classes.
For $l=4$ these are ${\bf c} = 000$ (+), ${\bf c} = 001$ (*), 
${\bf c} = 010$ (diamonds) and ${\bf c} = 111$ (triangles), while for the $l=8$ cases
they are ${\bf c} = 0000000 $ (+), ${\bf c} = 0000001 $ (*), 
${\bf c} = 0000010 $ (diamonds), ${\bf c} = 0000011 $ (triangles) and we have shown 
the roots associated with the remaining classes as small dots. 
The dashed circles correspond to $\| z \| =1$ and $\| z \| =r$ and have been 
inserted as a guide to the eye. All classes have a root near
$z = (1,0)$. The remaining roots cluster around and beyond the circle $\| z \| =r $. }
\label{rootplot}
\end{figure}

Considering the matching probability, we find
\begin{equation}
p(m,{\bf c}) = - \sum_{i=0}^{{\cal N}(\lambda)} \frac{1}{2\pi i} \oint_{\partial D_i} dz \frac{1}{z^{m+1}} \; \; \frac{1}{(1-z)\lambda(z;{\bf c})},  
\end{equation}
where $\partial D_i$ is a clock-wise contour around the $i^{\rm th}$ zero of 
$(1-z)\lambda(z;{\bf c})$, ${\cal N}(\lambda)$ is the number of zeroes, and we assume
that the zeroes are ordered such that 
$z_0 = 1 < z_1 < \| z_2 \| < \ldots < \| z_{{\cal N}} \|$. 
Evaluating explicitly the residues for the first two poles we have,
\begin{eqnarray}
p(m,{\bf c}) &=& 1 - A_1 \left ( \frac{1}{z_1} \right )^{m+1} \nonumber \\
&-& \sum_{i=2}^{{\cal N}(\lambda)} 
\frac{1}{2\pi i} \oint_{\partial D_i} dz \frac{1}{z^{m+1}} \; \; \frac{1}{(1-z)\lambda(z;{\bf c})}, 
\nonumber \\
\end{eqnarray}
with the residue $A_1$ given by
\begin{equation}
A_1 = \frac{1}{\lambda^{\prime} (z_1;{\bf c})( 1 - z_1)}.
\label{eqn:A1def}
\end{equation}
The remaining zeroes $z_2, z_3, \ldots z_l$ of $\lambda(z;{\bf c})$ 
are located near and beyond $\| z \| \approx r $, so that in the limit of large $m$, their 
relative contributions are smaller. We thus arrive at the asymptotic form
\begin{equation}
p(m,{\bf c}) \rightarrow 1 - A_1 \left ( \frac{1}{z_1} \right )^{m+1}   
\label{eqn:pmcasyexact}
\end{equation}
for large $m$. 

We can obtain approximate expressions for $z_1$ and thus an approximation of the
asymptotic behavior as follows. With 
\begin{equation}
\lambda(z;{\bf c}) = z^l + r^l(1-z) \left  [ 1 + c(z/r) \right ] 
\end{equation}
we see that when $\| z \| \sim 1$, the second term in the above equation is
a large term, $r^l$, multiplied with a term that will be small due to the 
$z-1$ prefactor. The product of these two terms can be made of order $1$, if 
$z-1 \sim 1/r^l$, which then can be made to cancel the first term $z^l$ if $z>1$.
Using the Lagrange Inversion Formula, $z-1$ can be expanded in a power series in $1/r^l$
 \cite{wilf}: Letting $u = z-1$ and $t = 1/r^l [1 + c(1/r)]^{-1}$, the equation 
$\lambda(z;{\bf c}) = 0$ can be written in the form 

\begin{equation}
u = t\; \phi(u),  
\end{equation}
where
\begin{equation}
\phi(u) = (1+u)^l \frac{1 + c \left( \frac{1}{r} \right ) }{1 + c \left(\frac{1+u}{r} \right )}.
\end{equation}
is a formal power series in u. Thus
\begin{equation}
z_1 = 1 + u(t) = 1 + \sum_{i=1}^{\infty} u_i t^i,  
\label{eqn:z1expan}
\end{equation}
with 
\begin{equation}
u_i = \frac{1}{i!} \left. \frac{{\rm d}^{i-1} \phi^i}{{\rm d}u^{i-1}}\right |_{u=0}. 
\end{equation}

One finds to leading and sub-leading order 
\begin{equation}
u_1 = 1,
\label{eqn:zeta1def}
\end{equation}
and
\begin{equation}
u_2 = l  - \frac{1}{r} \; \frac{ c^\prime(1/r)} { 1 + c(1/r)},
\end{equation} 
with
\begin{equation}
c^\prime(1/r) = \sum_{i=1}^{l-1} ic_i \left ( \frac{1}{r} \right )^{i-1}
\end{equation}
so that to leading order we have
\begin{equation}
z_1 = 1 + \frac{1}{1 + c(1/r) }\; \; \frac{1}{r^l}  
\label{eqn:z1leadingorder}
\end{equation}
The residue $A_1$ can be evaluated similarly, 
and we find to order $1/r^l$ that
\begin{equation}
A_1 = 1 - \frac{1}{r^l} \frac{ \frac{1}{r}c^\prime(1/r)} {\left [ 1 + c(1/r) \right ]^2} .
\label{eqn:A1expansion}
\end{equation}
Note that $A_1$ is of order one.

The asymptotic behavior of $h(b;{\bf c})$ and $d(b;{\bf c})$ can be worked out
in a similar manner. For large $b$ we find,
\begin{equation}
h(b) \rightarrow h_{asy}(b) = \frac{1}{r^l}\; \frac{A_1}{z_1} \left [ r^l \left ( z_1 - 1 \right ) \right ]^2 \; 
\left ( \frac{r}{z_1} \right )^b  
\label{eqn:hasy}
\end{equation}
and
\begin{equation}
d(b) \rightarrow d_{asy}(b) = \frac{A_1}{z_1} \left [ r^l \left ( z_1 - 1 \right ) \right ] 
\; \left ( \frac{r}{z_1} \right )^b.
\label{eqn:dasy}
\end{equation}
where $A_1$ is given by Eq.~(\ref{eqn:A1def}), $z_1$ is the smallest root of 
$\lambda(z;{\bf c})$ and from the expansion of $z_1$, Eq.~(\ref{eqn:z1expan}), we see that
the terms in square brackets are of order one

Taking the asymptotic forms of $h$ and $d$ to calculate the $n$-match
distribution one finds
\begin{equation}
p^{(1)}(n;m,{\bf c}) = 
A_1\left ( \begin{array}{c} m  + n \\ n \end{array} \right ) 
\;
\left [ A_1 r^l \left ( 1 - \frac{1}{z_1} \right )^2 \right ]^n \left ( \frac{1}{z_1} \right )^{m+1-n}.
\label{eqn:p1nm}
\end{equation}

\vspace*{1cm}
\begin{figure}[!ht]
\begin{center}
\includegraphics[width=10cm]{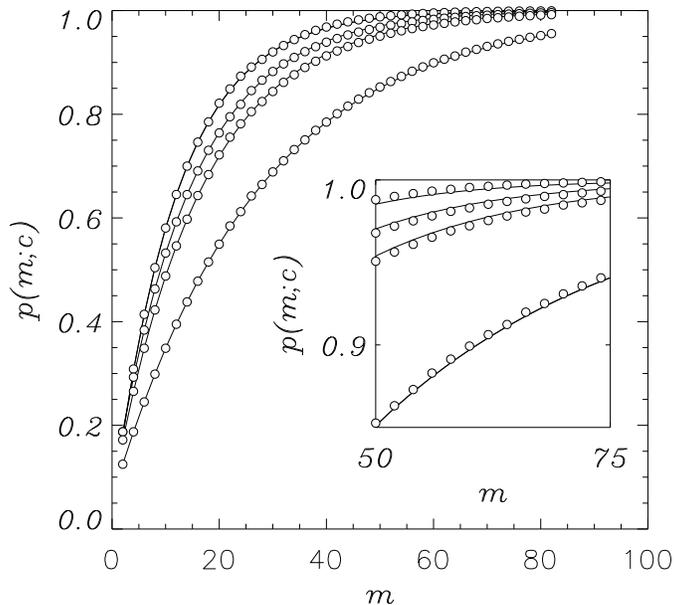}
\caption[]{Comparison of the asymptotic form, Eq.~(\ref{eqn:pmcasyexact}), with the
exact matching probabilities $p(m;\bf c)$. The figure shows
the matching probability for $l=4$, $r=2$, and $6 < k = m+l < 88$. The open circles
correspond to the numerically obtained matching probabilities for 
 ${\bf c} = 000$, $001$, $010$ and $111$ (from top to bottom). The lines 
correspond to the asymptotic form, Eq.~(\ref{eqn:pmcasyexact}), with $A_1$ and
$z_1$ calculated numerically. Inset: Plot of
$p(m;\bf c)$ for intermediate values of $m$. The symbols are as in the main figure. 
The equivalence classes are (from top to bottom): $000$, $001$, $010$ and $111$.      
}
\label{pklplot}
\end{center}
\end{figure}

Figure \ref{pklplot} shows a comparison of the asymptotic form, Eq.~(\ref{eqn:pmcasyexact}), 
with the exact matching probabilities $p(m;\bf c)$. The figure shows
the matching probability for $l=4$, $r=2$, and $6 < k = m+l < 104$. For $l=2$ 
there are $4$ equivalence classes: $000$, $001$, $010$, and $111$ with $6$, $6$, 
$2$ and $2$ members, respectively ({\em cf.} Tables \ref{bitv1} and \ref{bitv2} 
for the case  $r=2$). The open circles
correspond to the numerically obtained matching probabilities for 
${\bf c} = 000$, $001$, $010$ and $111$ (from top to bottom). For $m \le 20$ $(k \le 24)$, 
$p(m;{\bf c})$ was obtained by direct enumeration of all possible strings and checking
for matches. For $m > 20$ a sampling algorithm was used: for each value of $k$, $10^6$
strings of length $k$ were generated randomly and the matching probability was obtained 
by counting the matching strings of the sample. The solid lines 
correspond to the asymptotic form, Eq.~(\ref{eqn:pmcasyexact}), with $z_1$ and
$A_1$ calculated numerically, from Eqs.~(\ref{eqn:lambdazdef}) and (\ref{eqn:A1def}) for
each of the equivalence classes. The inset shows $p(m;\bf c)$ for intermediate values of $m$,
where we do not expect the asymptotic form to be very good.   

The discrepancies become much more severe when we consider the $n$-match distribution.
Fig. \ref{ndist_plot} shows the $n$-match distributions for a $4$ letter binary string
inside a random string of length $k=256$ for the four possible equivalence classes 
${\bf c} = 000$ (top left), ${\bf c} = 001$ (top right), ${\bf c} = 010$ (bottom left),
and ${\bf c} = 111$ (bottom right). The solid circles are the exact matching probabilities
that were obtained numerically using the algorithm described above. 
The dotted line corresponds to the  approximation Eq.~(\ref{eqn:p1nm}), normalized by an 
overall constant. The dashed line corresponds to the gaussian approximation of Kleffe 
and Borodovsky \cite{kleffe}, while the dot-dashed line is the compound poisson 
approximation of Chrysaphinou and Papastavridis \cite{chrys}, Geske {\it et al.} \cite{geske},  
and Schbath \cite{schbath}.
 
Note that while the 
approximation Eq.~(\ref{eqn:p1nm}) performs very poorly, the gaussian and compound-poisson 
distributions approximate well the true distribution only for some equivalence 
classes ${\bf c}$, but fail for others, as was noted by Robin and Schbath 
\cite{robschbath}. The solid line on the other hand, is the single analytical result of this 
article and agrees well with the actual distributions. 
We now turn to the description of the $n$-match probability in terms
of the (configurational) partition function of a 1d lattice gas.  

\vspace*{1cm}
\begin{figure}[!ht]
\begin{center}
\end{center}
\includegraphics[width=16cm]{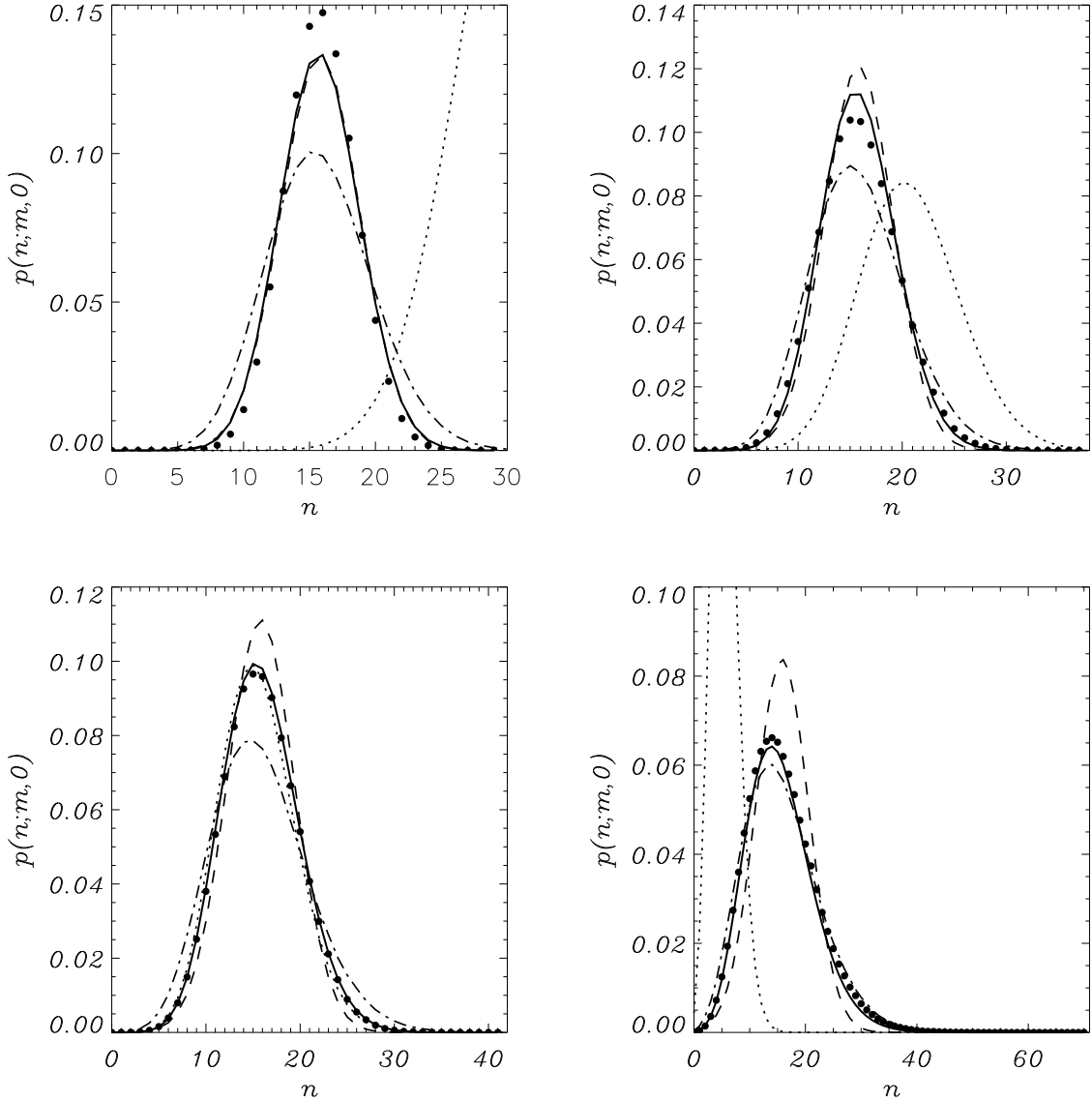}
\caption[]{The $n$-match distribution for matching a $l=4$ letter binary 
string $x$ inside a random string of length $k=256$, for $x=0001$ (top left),
$x=1001$ (top right), $x=1010$ (bottom left) and $x=1111$ (bottom right).
The circles are the exact probabilities, the dotted line corresponds to 
the approximation Eq.~(\ref{eqn:p1nm}) (normalized by an overall constant) and
the dashed and dashed-dotted lines correspond to the Gaussian and compound 
poisson approximation (see text for details). The solid line is the analytical 
result of this paper.
}
\label{ndist_plot}
\end{figure}

\section{The $n$-match probability as the partition function of a 1d lattice gas}
In this section we present the statistical mechanics approach to calculating the $n$-match
distribution function. We first map the problem into one of calculating the 
(configurational) partition sum of a $1d$-lattice gas. We next analyze the interaction 
emerging in such a description, then set up a virial expansion leading to an 
approximate evaluation of the partition function and finally discuss asymptotic limits. 

\subsection{The $n$-particle partition function}

Our starting point is Eq.(\ref{eqn:pnmaspartition}), which we reproduce below for 
convenience,
\begin{equation}
p(n;m,{\bf c})
= \sum_{a_1<a_2<\cdots<a_n} \frac{1}{r^{m+l}} d(a_1) 
\left [ \prod_{i=1}^{n-1}h(a_{i+1}-a_{i}) \right ] d(m-a_n).
\end{equation}
with $d$ and $h$ as defined in Eqs.~(\ref{eqn:ddef}) and (\ref{eqn:gdef}). 
The expression above for $p(n;m,{\bf c})$ already resembles 
the partition function of a gas of $n$ particles with particle boundary 
interactions proportional to $-\ln d$ and nearest neighbor particle-particle
interactions proportional to $- \ln h$. 
In order to make this analogy work, we need to consider what we mean by the 
free-particle, i.e. no interaction limit. 

Note that $d(b)$ and $h(b)$ are 
conditional matching weights. For example, 
$h(b)$ is the weight of the compound event: given a match at position $a$ 
what is the likelihood that the next 
match is at $a+b$. The asymptotic behavior of $d(b)$ and $h(b)$,  
Eqs.~(\ref{eqn:dasy}) and (\ref{eqn:hasy}), can be interpreted 
to correspond to the approximation when the correlations inherent in 
the compound events are ignored.  Thus the ratios 
$d(b)/d_{asy}(b)$ and $h(b)/h_{asy}(b)$ measure the strength of 
the correlations in such events.  

It is therefore natural to define the 
particle-boundary and particle-particle interactions,
$U^{boun}(b)$ and $U(b)$, respectively as
\begin{eqnarray}
e^{-\beta U^{boun}(b)} &=& \frac{d(b)}{d_{asy}(b)} \label{eqn:Ubdef} \\
e^{-\beta U(b)} &=& \frac{ h(b)}{h_{asy}(b)}. \label{eqn:Udef}
\end{eqnarray}
We thereby obtain meaningful physical interactions that 
vanish as $ b \rightarrow \infty $.   
Note that since the potentials do not have any characteristic
scale, a temperature by itself is meaningless and we will write 
"energies" always with the  pre-factor $\beta$, {\em i.e.} in dimension-less units. 

The (configurational) partition function, Eq.~(\ref{eqn:pnmaspartition}) 
can now be written in terms of these interactions as 
\begin{equation}
p(n;m,{\bf c}) = \frac{A_1}{z_1^{m+1}} e^{\beta \mu n} \;
\sum_{a_1<a_2<\cdots<a_n}e^{-\beta  \mathcal{H}_n(a_1,\ldots , a_n)  }, 
\label{eqn:pzmmdef_good}
\end{equation}
with
\begin{equation}
e^{\beta \mu } = A_1 \frac{r^l}{z_1}\;\left (z_1 -1 \right )^2
\label{eqn:ebmudef}
\end{equation}
and the Hamiltonian given by
\begin{equation}
\mathcal{H}_n(a_1,\ldots , a_n) = U^{boun}(a_1) + U^{boun}(m-a_n) + \sum_{i=1}^{n-1} U(a_{i+1} - a_i)
\label{eqn:Hamiltonian}
\end{equation}

Eq.~(\ref{eqn:p1nm}) corresponds to the free-particle limit ($U=U_b = 0$), 
which in the probability language is the limit of all correlations suppressed.
Before proceeding, it is instructive to study these interactions in more detail.

\subsection{Interactions}
\label{interac}
Consider the particle-particle interaction first. From Eqs.~(\ref{eqn:hasy}) and 
(\ref{eqn:Udef}) we find that 
\begin{equation}
e^{-\beta U(b)} = \left [ \frac{z_1}{A_1} \; \frac{1}{r^{2l} (z_1-1)^2} \right ] h(b) r^{l-b} z_1^b ,
\label{eqn:ppint}  
\end{equation}
where the term in square brackets is of order one, with respect to the small parameter 
$1/r^l$, cf. Eqs.~(\ref{eqn:A1expansion}) and (\ref{eqn:z1expan}). 

Figure \ref{ueffl0406} shows the particle particle interactions 
for words of length $l=4$ and $l=6$ as parameterized
by their associated equivalence classes ${\bf c}$. 
The potentials are plotted against distance measured in 
units of the word length $l$ and have been vertically 
offset for clarity with the dashed lines representing 
$U=0$. The crosses on the dashed line indicate
that the associated potential at that value is $+\infty$. 
We see that the potential have infinite values only
for $b \le l$. Also, the values of the potential in the regime $b \le l$ 
are generally much bigger than in the regime $b >l $, meaning that
the potential is stronger in the former region. We will refer to 
the region $b \le l$ and $b > l$ as the core and tail of the 
interaction, respectively. 
For a given length $l$ and depending on ${\bf c}$, we also see that 
the interactions have different features. For 
${\bf c} = 0\cdots0$ the interaction has a hard-core of size 
$l$ followed by a repulsive tail, while for ${\bf c} = 1\cdots1$
the interaction has a strongly attractive compontent at $b=1$, followed
by a hard-core region for $1 \le b \le l$, that goes over into 
an oscillatory but decaying 
tail. The potentials for the other values of ${\bf c}$ seem to be a 
mixture of these two types of behavior.

Figure \ref{ueff0011} shows the behavior of the potentials associated 
with the equivalence classes ${\bf c} = 0\cdots0$ (left) and  
${\bf c} = 1\cdots1$ (right) in their dependence on the word length $l$. 
For both equivalence classes we see
that the tail of the interaction becomes weaker as $l$ increases.
When ${\bf c} = 0\cdots0$, the core is hard-core and only
the core-size $l$ changes. The situation is different for 
${\bf c} = 1\cdots1$. For the ${\bf c} = 1\cdots1$ family of interactions we
see that the attractive part of the core actually becomes stronger with 
increasing $l$. It turns out that the same is also true 
for the other equivalence classes, namely with increasing $l$, 
the cores of the interactions become stronger, while the tails 
become  weaker.

\vspace*{1cm}
\begin{figure}[!ht]
\begin{center}
\end{center}
\includegraphics[width=16cm]{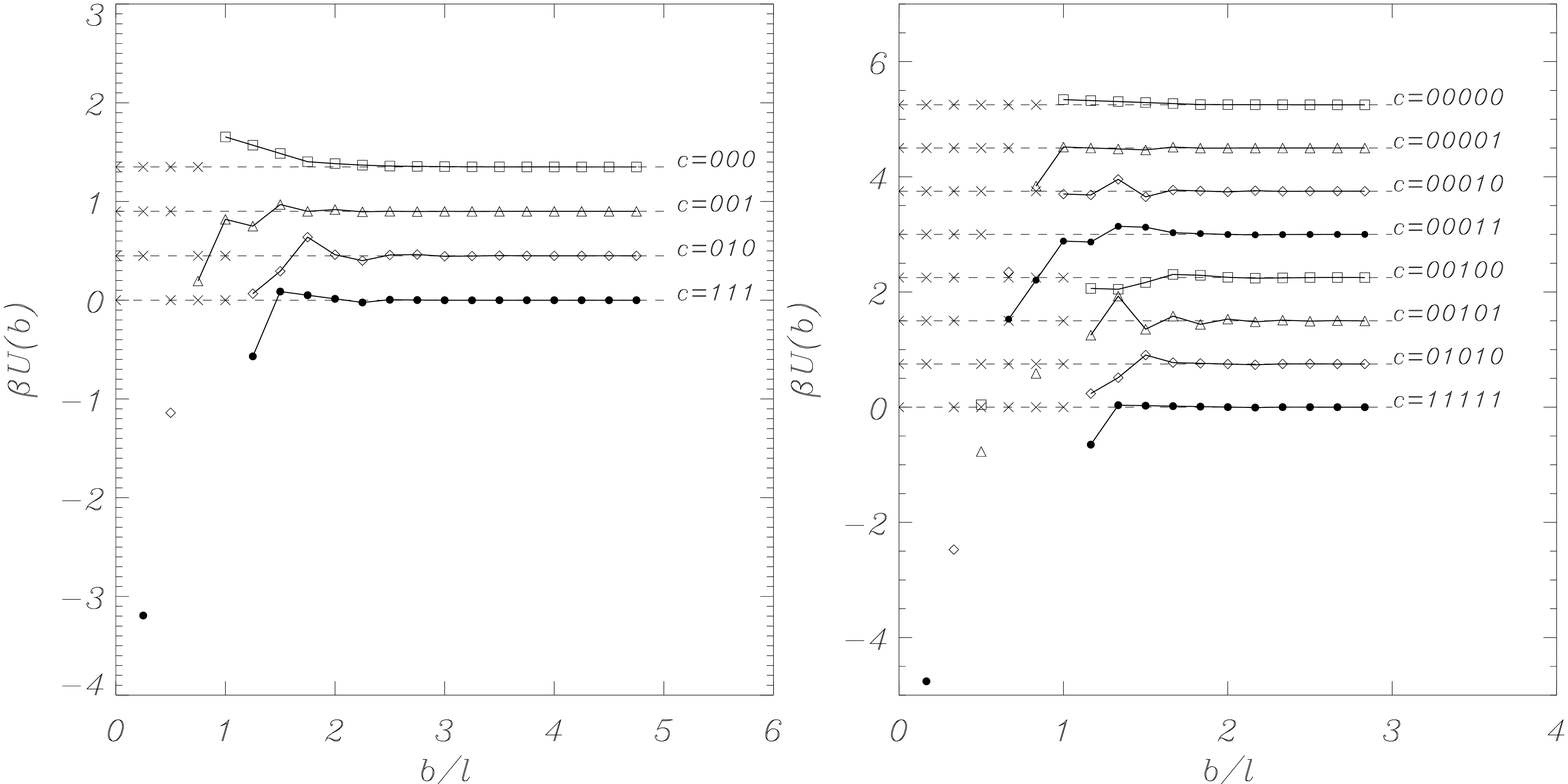}
\caption[]{Plot of the effective potentials $\beta U(b)$, 
Eq.~(\ref{eqn:ppint}), associated with the equivalence classes
${\bf c}$ of strings of lengths $l$. The potentials are 
plotted against distance measured in units of the word length $l$.
Note that the potentials have been vertically 
offset for clarity. The dashed lines represent the 
$U=0$ lines for each potential. The crosses on the line U=0 indicate
that the associated potential at that point is $+\infty$.  
Left: Interparticle potentials associated with words of length $l=4$, for
which the possible equivalence classes are ${\bf c} = 000, 010$ and
$111$, as indicated in the figure. Right: same as left but for $l=6$.
Notice how the attractive part of the interaction emerges and grows
stronger as the fundamental period of the string decreases to $1$ (
${\bf c} = 1\cdots1$). The tail of the interaction corresponds to 
the regime $b/l >1$.
}
\label{ueffl0406}
\end{figure}

In summary, Figs. \ref{ueffl0406} and \ref{ueff0011} suggest 
the following generic features of the interactions:
(i) a strong core $b \le l$, followed by a weak tail for $b > l$, 
and, (ii) for a given family of interactions, as $l$ increases the core 
of the interaction tends to become stronger, while the tail of the 
interaction becomes weaker. 

\vspace*{1cm}
\begin{figure}[!ht]
\begin{center}
\end{center}
\includegraphics[width=16cm]{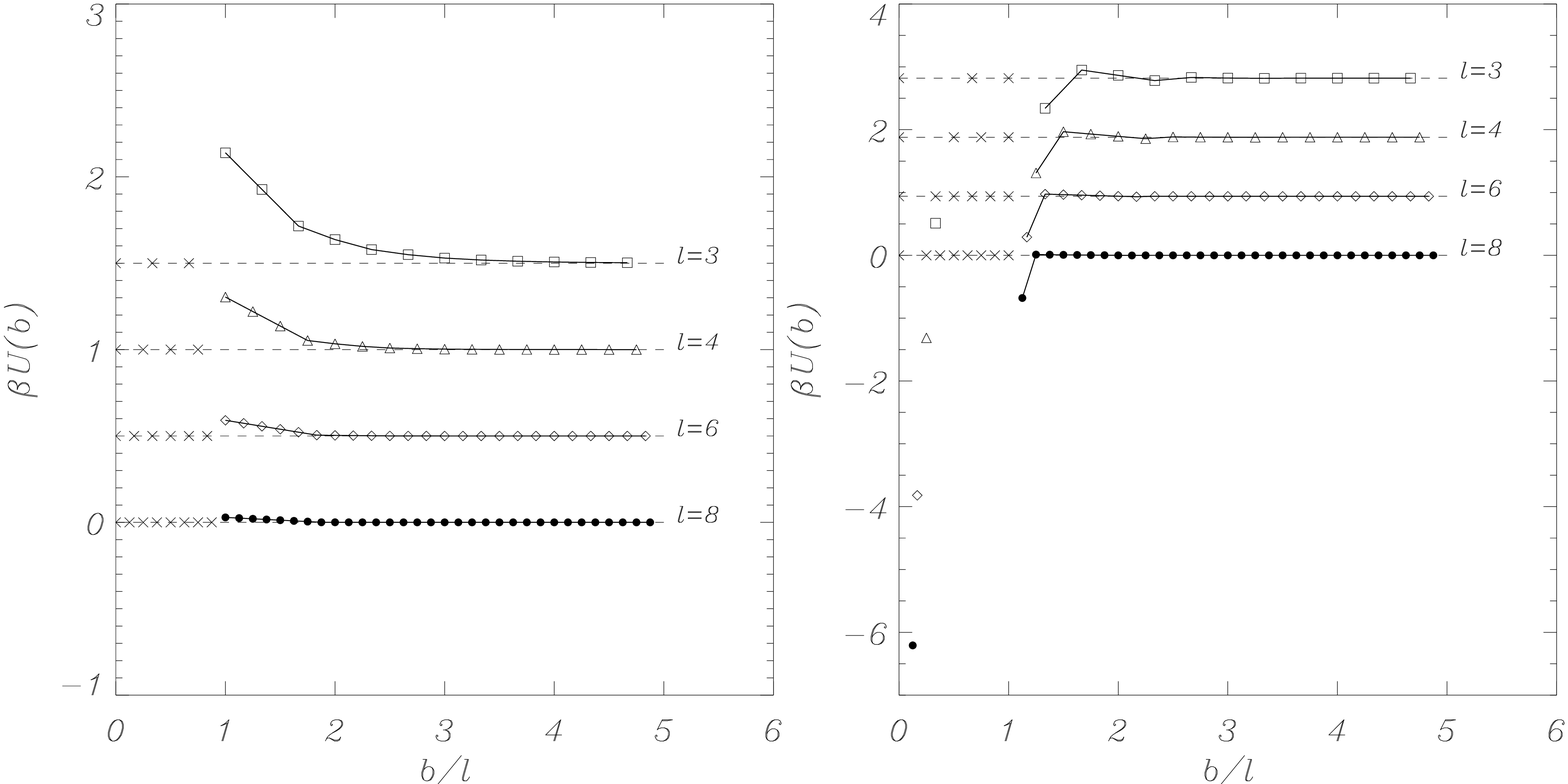}
\caption[]{Plot of the effective potentials $\beta U(b)$, 
Eq.~(\ref{eqn:ppint}), associated with the equivalence classes
${\bf c} = 0\cdots0$ and $1\cdots1$ and their dependence on
the lengths $l$. The potentials are 
plotted against distance measured in units of the word length $l$.
Note that the potentials have been vertically 
offset for clarity. The dashed linerepresent the 
$U=0$ line for each potential. The crosses on the a line indicate
that the associated potential at that value is $+\infty$.  
Left: Interparticle potentials associated with the equivalence 
class ${\bf c} = 0\cdots0$ for words of length $l=3,4,6$ and $8$.
Note that the interactions have a hard-core of size $b/l =1$ followed
by a repulsive tail. The strength of the tail weakens with increasing
$l$. Right: Interparticle potentials associated with the equivalence 
class ${\bf c} = 1\cdots1$ for words of length $l=3,4,6$ and $8$.
Note that the interactions have an attractive part at $b=1$, followed
by a hard-core for $b/l < 1$, and 
a weak, oscillatory decaying tail. Also note the opposite behavior
of the strength of the core and the tail: With increasing $l$, the strength 
of the attractive part of the core is seen to increase, 
while the strength of the tail 
decreases.
}
\label{ueff0011}
\end{figure}

These observations can be readily proven from the small $b$ behavior 
of $h(b)$, which in turn can be extracted from the recursion for $h(b)$, 
Eqs.~(\ref{eqn:grec}) and (\ref{eqn:Concemore}}). Thus we find
for $b < l$ 
\begin{equation}
h(b) =  \left \{ \begin{array}{ll} 
            c_b, & \mbox {if $\chi$ does not divide $b$,} \\ 
            1, & \mbox{if $b = \chi $,} \\
            0, & \mbox{otherwise,} \\
\end{array} \right.
\label{eqn:hcore} 
\end{equation}
where $\chi$ is the fundamental period associated with ${\bf c}$ that was 
defined at the end of Section \ref{bit-sect}. Recall that by definition,
$h(0) = 0$. 

Thus the interaction in the core region can be written as 
\begin{equation}
\beta U(b) =  -\ln h(b) \; + b \ln \left ( \frac{r}{ z_1} \right ) \; -l \ln r \; + \beta U_0,  \; \; \; \mbox{$b \le l$} 
\label{eqn:Uppcore}
\end{equation}
where
\begin{equation}
\beta U_0 = \ln \left ( \frac{z_1}{A_1 \left [r^l (z_1-1) \right ]^2} \right )
\end{equation}
is a constant that is of order $1/r^l$, since the argument of the logarithm
is of order 1 to the same order.

We see that the interaction becomes $+\infty$, whenever $h(b) = 0$. This is
certainly the case for $b < \chi$. Furthermore, since $r/z_1 > 1$, in the core 
region finite values of  $U(b)$  increase with increasing $b$, as clearly seen 
in Fig. \ref{ueffl0406}. 

The first finite value of $U(b)$ occurs at $b = \chi$. 
From Eq.~(\ref{eqn:Uppcore}) we obtain for $U(\chi)$
\begin{equation}
\beta U(\chi) = \chi \ln \left ( \frac{r}{z_1} \right ) - l \ln r + \mathcal{O} \left (\frac{1}{r^l} \right ). 
\label{eqn:Uxhi}
\end{equation}
Thus it is apparent that for fixed $\chi$,  $\beta U(\chi)$ becomes more negative as either
$l$ or $r$ increase. In fact we see that to leading order, the dependence of  
$\beta U(\chi)$ on $l$ is linear, while its dependence on $r$ is logarithmic.
Also for $\chi < l$ we see that $\beta U(\chi)$  is negative.

The case when $\chi = l$, corresponding to ${\bf c} = 00\ldots0$, 
is a little more complicated. In that case the $\ln r$ terms in 
Eq.~(\ref{eqn:Uxhi}) cancel and we are left with a term $l \ln z_1$ which is of order
$1/r^l$ as well and thus $\beta U_0$ cannot be neglected anymore. However this means
that the potential is of order $1/r^l$, which turns out to be the correct scale of the
strength of the tail and indeed decreases as $r$ or $l$ increase (see Fig.\ref{ueff0011}).
To be specific, for $\chi = l$, it can be readily checked
that $h(b) = r^{b-l}$ for $l \le b < 2l$ and the potential in this regime thus becomes 
\begin{equation}
\beta U(b) = \frac{2l + 1 - b}{r^l} + \mathcal{O} \left (\frac{1}{r^{2l}} \right ), 
\end{equation}
where we have substituted the expansions of $A_1$ and $z_1$, Eqs.~(\ref{eqn:A1expansion})
and (\ref{eqn:z1expan}), respectively, to lowest non-trivial order. Thus we see that 
for $\chi = l$ the characteristic energy scale of the tail of the 
interaction scales like  $\sim l/r^l$, and 
decreases as $l$ or $r$ increase.
 
The case of general $\chi$ and ${\bf c}$ is similar, but the calculation are tedious 
yet straightforward. Rather than doing this, we will motivate the result by 
considering the value of the interaction at $b = l + \chi$, which is readily worked out from
\begin{equation}
h(l+\chi) = \left \{ \begin{array}{ll} 
            r^\chi - h(l) - 1, & \mbox {for $\chi < l/2$,} \\ 
            r^\chi - h(l) - 1, - \sum_{\tau = \chi + 1}^{l-1} h(\tau) c(l+\chi-1), 
	              & \mbox{for  $l/2 \le \chi < l $,} \\
            r^\chi - h(l), & \mbox{for  $\chi = l $.} 
\end{array} \right.
\end{equation}
This means that 
\begin{equation}
h(l+\chi) = r^\chi \left ( 1 - \epsilon \right ),
\end{equation}
where $\epsilon r^\chi$ is at most $l-\chi+1$ and hence of order $l$. 
Substituting this result into 
the expression for $U(b)$ along with the expansions for $z_1$ and $A_1$, one finds that 
the result is of the form
\begin{equation}
-\beta U(l+\chi) =  \frac{\alpha}{r^l} + \mathcal{O} \left (\frac{1}{r^{2l}} \right ), 
\end{equation}
where $\alpha$ is a ${\bf c}$ and $l$ dependent constant of order one. 

To conclude, we find that the characteristic energy of the core of the interaction
scales like $-(l-\chi) \ln r$ ($\chi < l$), while the energy of the tail goes to leading 
order like $1/r^l$. These results are consistent with the behavior observed in 
Figs.~\ref{ueffl0406} and \ref{ueff0011}   

Turning to the particle boundary interactions, note that Eq.~(\ref{eqn:drec}), 
which can be conveniently written as
\begin{equation}
\frac{d(b)}{r^b} = 1 - \sum_{a=1}^{b} \frac{h(a)}{r^a}
\end{equation}
relates the properties of $h$ to those of $d$. We thus see that
analogous results can be obtained for the boundary interaction $U^{boun}(b)$ and
we leave the details to the interested reader.
   
\subsection{The Hamiltonian} 

The results of the previous section allow us to obtain approximate expressions for the 
probability $p(n;m,{\bf c})$, by first approximating the effective Hamiltonian
$H_n$ and then carrying out the configurational sums. This is most easily done
using generating functions.

Define the generating functions associated with Eqs.~(\ref{eqn:Ubdef}) 
and (\ref{eqn:Udef}) as
\begin{eqnarray}
D(z) &=& \sum_{b=0}^{\infty} z^b e^{-\beta U_b(b)},  \\
H(z) &=& \sum_{b=0}^{\infty} z^b e^{-\beta U(b)} .
\end{eqnarray}
It is not difficult to show that in terms of the generating functions of $d(b)$ and
$h(b)$, $D(z)$ and $H(z)$ are given by
\begin{eqnarray}
D(z) &=& e^{-\beta \mu} (z_1-1) d\left (\frac{zz_1}{r};{\bf c} \right ) \label{eqn:Ddef} \\
H(z) &=& e^{-\beta \mu} h\left (\frac{zz_1}{r};{\bf c} \right ) \label{eqn:Hdef}.
\end{eqnarray}
Using the convolution property, Eq.~(\ref{eqn:pzmmdef_good}) can be
written in terms of the generating functions $D(z)$ and $H(z)$ as
\begin{equation}
p(n;m,{\bf c}) = \frac{A_1}{z_1^{m+1}} e^{\beta \mu n}  
\frac{1}{2\pi i} \oint_{\partial D} dz \frac{1}{z^{m+1}} \; \; D^2(z) H^{n-1}(z),
\label{eqn:pnmc_contour}
\end{equation}
where the contour is again the boundary of a domain enclosing the origin inside of 
which $D^2(z)H^{n-1}(z)$ is analytic. Eq.~(\ref{eqn:pnmc_contour}) is the lattice 
analog of the partition function of a 1d gas with pairwise nearest neighbor interactions.
The 1d continuum case has been treated in detail by G\"ursey \cite{gursey} 
({\em see also} Fisher \cite{fisher}). 
 
Next, define the truncated generated functions $D_{\Lambda}(z)$ and $H_{\Lambda}(z)$
as
\begin{eqnarray}
D_{\Lambda}(z) &=& \sum_{b=0}^{\Lambda -1} z^b e^{-\beta U_b(b)}, \label{eqn:Ubzdef} \\
H_{\Lambda}(z) &=& \sum_{b=0}^{\Lambda -1} z^b e^{-\beta U(b)} \label{eqn:Uzdef}.
\end{eqnarray}
It is readily seen that these generating functions are associated with the Boltzmann
factor of an interaction that has been cut-off at $b \ge \Lambda$. The idea is that
since, by construction, the interactions decay to zero at large distances, introducing
a finite cut-off $\Lambda$ will introduce only a small and controllable error in the 
overall calculation. In what follows, we will use this to set up a perturbation expansion
of the probability distribution. 
We need to note however that since the result has to be a normalized distribution, setting
the potential to zero beyond the cut-off will destroy the normalization of the distribution.
Indeed there are at least two ways to handle the interaction beyond the cut-off:
(i) we can either set the interaction to a  constant $U_\Lambda$ for $b \ge \Lambda$ and 
eventually choose $U_\Lambda$ such that the distribution is normalized, or (ii) we take the 
interaction beyond $\Lambda$ to be rapidly decaying. It turns out that the calculation 
can be done for either of the cases. The approximation by a constant potential beyond the 
cut-off lends itself readily for obtaining error bounds, as we will sketch below.
On the other hand, it turns out that the tail of the actual interactions 
{\em does} asymptotically decay exponentially. 
Thus letting the interaction decay exponentially beyond the cut-off turns out to be a 
very good approximation and we will calculate the probability distributions in this way.

Consider the case of a constant potential beyond the cut-off first and  
define the approximate interaction $\hat{U}(b)$ as
\begin{equation}
\hat{U}_\Lambda(b) = \left \{ \begin{array}{ll}  U(b), & b < \Lambda \ \\
                                      U_\Lambda,   & b \ge \Lambda, \end{array} \right.
\end{equation}
with the corresponding generating function given by
\begin{equation}
\hat{H}_\Lambda(z) = \sum_{b=0}^{\infty} z^b e^{-\beta \hat{U}_\Lambda(b)} = H_{\Lambda}(z) 
+ e^{-\beta U_\Lambda} \; \frac{z^\Lambda}{1-z}.
\end{equation}

Since $d(z)$ is related to $h(z)$ via Eq.~(\ref{eqn:dzexplicit}), this implies 
a corresponding boundary interaction which can be worked out as
\begin{equation}
\hat{D}_\Lambda(z) = \sum_{b=0}^{\infty} z^b e^{-\beta \hat{U}_\Lambda(b)} = D_{\Lambda}(z) 
+ e^{-\beta U_\Lambda} \; \frac{z^\Lambda}{1-z}.
\end{equation}

Define the approximation to $p(n;m,{\bf c})$, Eq.~(\ref{eqn:pnmc_contour}), as
\begin{equation}
\hat{p}(n;U_\Lambda,m,{\bf c}) = \frac{A_1 e^{\beta \mu n}}{z_1^{m+1}}   
\frac{1}{2\pi i} \oint_{\partial D} dz \frac{1}{z^{m+1}} \; \; 
\hat{D}_\Lambda^2(z) \hat{H}_\Lambda^{n-1}(z),
\end{equation}

It is clear that  as $\Lambda \rightarrow \infty$ we must have $U_\Lambda \rightarrow 0$,
since an increasingly larger part of the true interactions is kept. 
 
By using the definition of $U_\Lambda(b)$ and writing
$\hat{p}(n;U_\Lambda,m,{\bf c})$ in the partition sum form of 
Eq.~(\ref{eqn:pnmaspartition}), it can readily be verified that if 
\begin{equation}
U_{-} \le U_\Lambda \le U_{+}
\end{equation}
this implies that
\begin{equation}
\hat{U}_{-}(b) \le \hat{U}_\Lambda(b) \le \hat{U}_{+}(b)
\end{equation}
for all values of $b$, which in turn implies that
\begin{equation}
\hat{p}(n;U_{+},m,{\bf c}) \le  \hat{p}(n;U_\Lambda,m,{\bf c}) \le \hat{p}(n;U_{-},m,{\bf c}).
\end{equation}
Thus by choosing $U_{+}$ and $U_{-}$ as
\begin{eqnarray}
U_{+} &=& \max_{b \ge \Lambda} \left \{ U(b), U^{boun}(b) \right \}, \\
U_{-} &=& \min_{b \ge \Lambda} \left \{ U(b), U^{boun}(b) \right \}
\end{eqnarray}
one could in principle obtain error bounds on the approximate distribution, which will become
tighter as $\Lambda \rightarrow \infty$. We will not pursue this any further in the present 
article, but instead perform the calculation with an exponentially decaying interaction beyond
the cut-off $\Lambda$. 

Recall that the tail of the true interaction is due to the other 
zeroes of $\lambda(z;{\bf c})$, which
are located a distance $\sim r$ from the origin, ({\em see} Fig.\ref{rootplot}). 
Thus superposed on the asymptotic behavior 
of $h(b)$, which we have shown to fall-off like $z_1^{-b}$, there will be terms that 
decay more rapidly and roughly as $r^{-b}$, since $z_1 < r$. 
In fact it is the latter that are responsible for the asymptotic behavior 
of the interactions. For $b$ large, we therefore take approximately  
\begin{equation}
h(b) \approx e^{\beta \mu} z_1^{-b} + \gamma e^{\beta \mu} r^{-b} 
\end{equation}
which upon taking logarithms and factoring out the first terms implies that 
asymptotically
\begin{equation}
\beta U(b) \approx \beta \mu b \ln z_1 - \gamma \left ( \frac{z_1}{r} \right )^{b},  
\end{equation}
where we have neglected higher order terms $\gamma^k (z_1/r)^{kb}$. Of course,
with increasing cut-off $\Lambda$, the residual tail will be less important. 

This suggest taking the following approximate interactions:
\begin{equation}
\hat{U}_\Lambda(b) = \left \{ \begin{array}{ll}  U(b), & b < \Lambda \ \\
 -\gamma \left ( \frac{z_1}{r} \right )^{b} ,   & b \ge \Lambda, \end{array} \right.
\end{equation}
with the corresponding approximate generating function given by
\begin{equation}
\hat{H}_\Lambda(z) = \sum_{b=0}^{\infty} z^b e^{-\beta \hat{U}_\Lambda(b)} = H_{\Lambda}(z) 
+ \gamma  \left ( \frac{z_1}{r} \right )^{\Lambda}\; \frac{z^\Lambda}{1-\frac{z_1}{r}z}
+ \frac{z^\Lambda}{1-z}.
\label{eqn:hhat}
\end{equation}

Since $d(z)$ is related to $h(z)$ via Eq.~(\ref{eqn:dzexplicit}), this implies 
a corresponding approximate interaction for the boundary interaction, 
which can be worked out,
\begin{equation}
\hat{D}_\Lambda(z) = \sum_{b=0}^{\infty} z^b e^{-\beta \hat{U}_\Lambda(b)} = D_{\Lambda}(z) 
+ \gamma  \left ( \frac{z_1}{r} \right )^{\Lambda}\; \frac{z^\Lambda}{1-\frac{z_1}{r}z}
+ \frac{z^\Lambda}{1-z}.
\label{eqn:dhat}
\end{equation}
Denoting the generating function of the approximate tail of the interaction as
\begin{equation}
\Gamma(z) = \gamma  \left ( \frac{z_1}{r} \right )^{\Lambda}\; 
\frac{z^\Lambda}{1-\frac{z_1}{r}z},
\end{equation}
$\hat{p}(n;m,{\bf c})$ becomes
\begin{equation}
\hat{p}(n;\gamma,m,{\bf c}) = \frac{A_1 e^{\beta \mu n}}{z_1^{m+1}}   
\frac{1}{2\pi i} \oint_{\partial D} dz \frac{1}{z^{m+1}} \; \; 
\hat{D}_\Lambda^2(z) \hat{H}_\Lambda^{n-1}(z).
\label{eqn:pnmc_contour_app}
\end{equation}

What therefore remains to be done is to evaluate the contour integral, 
Eq.~(\ref{eqn:pnmc_contour_app}), which can be carried out by the method of 
stationary phase, which in the context of generating functions is also known as 
Hayman's method \cite{wilf}:

\subsection{Distributions}
Write the integral in Eq.~(\ref{eqn:pnmc_contour_app}) as
\begin{equation}
I =  \frac{1}{2\pi i} \oint_{\partial D} dz \frac{1}{z^{m+1}} \; \; f(z).
\end{equation}
Then for large $m$, the value of the integral is given approximately by
\begin{equation}
I \approx \left (\frac{1}{u_m}\right )^m \frac{f(u_m)}{\sqrt{2\pi b_m}},
\label{eqn:Iapp}
\end{equation}
where $u_m$ is the smallest positive real root of the equation
\begin{equation}
m = u  \frac{\rm d}{\rm du}\ln f(u)
\end{equation}
and $b_m$ is given by
\begin{equation}
b_m = u  \frac{\rm d} {\rm du} \ln f(u)+ u^2  \frac{{\rm d}^2} {{\rm du}^2} \ln f(u).
\end{equation}

Applying Hayman's method to the integral, Eq.~(\ref{eqn:pnmc_contour_app}), we let
\begin{equation}
f(u) =  \hat{D}_\Lambda^2(u) \hat{H}_\Lambda^{n-1}(u) 
\label{eqn:fudef}
\end{equation}
and find after a little bit of algebra
\begin{eqnarray}
m &=& u  \frac{{\rm d} }{{\rm du}} \ln f(u)  \nonumber \\ &=& \frac{2}{x}  
\frac{1 + \Lambda x + x^2 (1+x)^{\Lambda-2} \left [ \hat{D}^\prime_\Lambda\left (\frac{1}{1+x} \right ) 
+ \Gamma^\prime\left (\frac{1}{1+x} \right )  \right ]}
{ 1 + x (1+x)^{\Lambda-1} \left [ \hat{D}_\Lambda\left (\frac{1}{1+x} \right ) 
+ \Gamma\left (\frac{1}{1+x} \right )\right ] } \nonumber \\
&+& \frac{n-1}{x}  
\frac{1 + \Lambda x + x^2 (1+x)^{\Lambda-2} \left [ \hat{H}^\prime_\Lambda\left (\frac{1}{1+x} \right ) 
+ \Gamma^\prime\left (\frac{1}{1+x} \right )  \right ]}
{ 1 + x (1+x)^{\Lambda-1} \left [ \hat{H}_\Lambda\left (\frac{1}{1+x} \right ) 
+ \Gamma\left (\frac{1}{1+x} \right )\right ], } 
\label{eqn:mofx}
\end{eqnarray}
where we have parameterized $u$ as 
\begin{equation}
u = \frac{1}{1+x}.
\label{eqn:xu}
\end{equation}

Since we are interested in solutions for large $m$, it is clear from the above that to leading
order $x \propto 1/m$. Multiplying both sides of the above equation by $x$ and expanding 
the fractions in a power series around $x=0$, we obtain 
\begin{equation} 
mx = (n+1) \left \{ 1 + \epsilon_1 x + \epsilon_2 x^2 + \ldots \right \}
\label{eqn:mxeq}
\end{equation}
The first two orders can be readily worked out, yielding
\begin{equation}
\epsilon_1 = \Lambda - \tilde{\gamma}\xi -   
\frac{2\hat{D}_\Lambda(1) + (n-1)\hat{H}_\Lambda(1) } {n+1}
\label{eqn:eps1}
\end{equation}
and
\begin{eqnarray}
\epsilon_2 &=& \tilde{\gamma}\xi \left [ 2 \xi + \tilde{\gamma}\xi - 1 \right ]  
+ \frac{1}{n+1} \left \{ 2\hat{D}^2_\Lambda(1) + (n-1)  \hat{H}^2_\Lambda(1)
 + 2 \left [ 2 \hat{D}^\prime_\Lambda(1) + (n-1) \hat{H}^\prime_\Lambda(1) \right ] \right. \nonumber \\
&-& \left [ 2(\Lambda - \tilde{\gamma}\xi) -1 \right ] 
  \left. \left [ 2 \hat{D}_\Lambda(1) - (n-1) \hat{H}_\Lambda(1) \right ] \right \}, 
\end{eqnarray}
where 
\begin{equation}
\xi = \frac{1}{1-\frac{z_1}{r}}
\end{equation}
and $\tilde{\gamma} = \gamma (1-1/\xi)^\Lambda$.

Rewriting Eq.~(\ref{eqn:mxeq}) in a form suitable for Lagrange's Inversion Formula,
\begin{equation}
x = \frac{n+1}{m - (n+1)\epsilon_1} \left \{ 1 + \epsilon_2 x^2 + \ldots \right \},
\end{equation}
We obtain an expansion of $x$ in terms of $(n+1)/[m-(n-1)\epsilon_1]$ and the coefficients $\epsilon_i$ as 
\begin{equation}
x = \frac{n+1}{m-(n-1)\epsilon_1} + \left [ \frac{n+1}{m-(n-1)\epsilon_1} \right ]^3 \epsilon_2 \ldots \; .
\label{eqn:virialx}
\end{equation}
The term $b_m$ can be worked out in a similar manner and we find
\begin{equation}
b_m = m + \frac{n+1}{x^2} - (n+1)(\epsilon_1 + \epsilon_2) + \ldots \; , 
\label{eqn:bmexp}
\end{equation}
where the omitted terms are of order $x$ and higher.

Combining Eqs.~(\ref{eqn:pnmc_contour_app}), (\ref{eqn:Iapp}), (\ref{eqn:fudef}), with 
Eqs.~(\ref{eqn:virialx}) and (\ref{eqn:bmexp}) we finally obtain
\begin{equation}
\hat{p}(n;\gamma,m,{\bf c}) \approx \frac{A_1 e^{\beta \mu n}}{z_1^{m+1}}   
\left ( 1 + x \right )^m  \hat{D}_\Lambda^2 \left ( \frac{1}{1+x} \right ) 
\hat{H}_\Lambda^{n-1} \left ( \frac{1}{1+x} \right ) \frac{1}{\sqrt{2\pi b_m}}.
\label{eqn:phat}
\end{equation}

The strength of the tail, $\tilde{\gamma} $, is still undertermined and we will determine
it by fitting the approximate tail to the actual interaction in the interval 
$b \in [ \Lambda, \Lambda + l-1 ] $. Note that this way there are no adjustable 
parameters and since the tail is only approximate, the normalization is not 
perfect and is found to vary by a few percent. Alternatively, one can choose 
$\tilde{\gamma}$ such that normalization is achieved. In either of the cases
the distributions do not vary significantly, meaning that for a certain range of
$\tilde{\gamma}$ values, the shape of the distribution is robust. 

The solid lines in Fig.~\ref{ndist_plot} show the approximate distribution, 
Eq.~(\ref{eqn:phat}), for the four equivalence classes associated with words
of length $l=4$ and with $r=2$, $k=256$. We will refer to this approximation 
as the liquid theory approximation. In this and all the other 
results that we will present, the cut-off $\Lambda$ was chosen as $\Lambda = 3l$ 
and $x$ was expanded to 2nd order. The dashed lines in Fig.~\ref{ndist_plot} 
are the Gaussian approximation of Kleffe and Borodovsky (KB) \cite{kleffe} 
with the distribution mean and variance given by Eqs.~(\ref{eqn:nave}) 
and (\ref{eqn:sigman}). The dot-dashed lines are the compound poisson (CP) 
approximation of Chrysaphinou and Papastavridis \cite{chrys}, 
Geske {\it et al.} \cite{geske} and Schbath \cite{schbath}. 

The variation between actual and approximate
distributions, $p(n)$ and $\hat{p}(n)$, can be quantified by the 
{\it total variational distance} \cite{barbour} 
between the two distributions and is defined as
\begin{equation}
d_{TV}(p,\hat{p}) = \frac{1}{2}\sum_n \| \hat{p}(n) - p(n) \| .
\end{equation}
Table 3 shows the variational distances between the actual and approximate 
distributions depicted in Fig. \ref{ndist_plot}, $(l=4)$ and $k=256$.

\begin{table}
\begin{center}
\vskip 0.25cm
\begin{tabular}{||r|l|l|l|l||}
\hline
${\bf c}$  & $d^{L}_{TV}$ & $d^{NL}_{TV}$ & $d^{CP}_{TV}$ & $d^{KB}_{TV}$ \\
\hline
     000 &  0.052 &  0.053 &  0.189 &  0.052 \\ 
     001 &  0.035 &  0.031 &  0.079 &  0.075 \\ 
     010 &  0.011 &  0.003 &  0.108 &  0.071 \\ 
     111 &  0.032 &  0.021 &  0.047 &  0.148 \\ 
\hline
\end{tabular}
\caption{Total variational distance between the actual distribution and the various 
approximate distribution for the case $r=2$, $k=256$: 
liquid theory approximation (L), Eq.~(\ref{eqn:phat}),
the liquid theory approximation normalized by an overall constant (NL), the compound
poisson approximation (CP) and the gaussian approximation (KB). }   
\end{center}
\label{dtvl4_r2}
\end{table}
We see that the (un-normalized) liquid theory approximation, Eq.~(\ref{eqn:phat}) (L), 
as well as the liquid theory approximation normalized by an overall constant (NL) 
perform better then the compound poisson (CP) and gaussian approximation (KB). Note 
that for ${\bf c} = 000$, none of the approximations captures the 
height of the peak of the distribution accurately and we will remark on this shortly.  

Tables 4 and 5 show the total variational distances 
between the actual and approximate distributions for word lengths $l=3,4,5,6,7$ and
$l=8$ and string lengths $k$ chosen such that $k/r^l = 16$, {\it. i.e.} the 
distributions have approximately the same mean. Overall, the 
liquid theory approximation, Eq.~(\ref{eqn:phat}) (L) , as well as the  liquid theory 
approximation normalized by an overall constant (NL) perform better then or as well as 
the compound poisson (CP) and gaussian approximation (KB) taken by themselves. 
The CP approximation gives a better approximation for 
${\bf c} = 11\cdots 1$ and for some of the low and high  $\chi$
equivalence classes associated with $l=7$ and $l=8$. Also note that 
for $l \ge 6$ the CP approximation performs generally better than the KB 
approximation, as was noted before by Robin and Schbath \cite{robschbath}. 
The poor performance of the liquid theory approximation for the 
case $l=3$ and ${\bf c} = 00$ turns out to be due to the fact 
that the expansion of $x$ and $b_m$ 
to second order is not adequate. Upon calculating $x$ (and $b_m$) more accurately, 
the agreement with the actual distributions turns out to be nearly perfect.
 
\begin{table}
\begin{center}
\vskip 0.25cm
\begin{tabular}{||r|l|l|l|l||}
\hline
${\bf c}$  & $d^{L}_{TV}$ & $d^{NL}_{TV}$ & $d^{CP}_{TV}$ & $d^{KB}_{TV}$ \\
\hline
      00 &  ***** &  (0.933) &  0.227 &  0.006 \\ 
      01 &  0.027 &  0.008 &  0.156 &  0.084 \\ 
      11 &  0.018 &  0.016 &  0.121 &  0.131 \\ 
\hline
     000 &  0.052 &  0.053 &  0.189 &  0.052 \\ 
     001 &  0.035 &  0.031 &  0.079 &  0.075 \\ 
     010 &  0.011 &  0.003 &  0.108 &  0.071 \\ 
     111 &  0.032 &  0.021 &  0.047 &  0.148 \\ 
\hline
    0000 &  0.009 &  0.010 &  0.090 &  0.018 \\ 
    0001 &  0.018 &  0.016 &  0.056 &  0.043 \\ 
    0010 &  0.010 &  0.008 &  0.061 &  0.050 \\ 
    0011 &  0.040 &  0.036 &  0.034 &  0.089 \\ 
    0101 &  0.021 &  0.024 &  0.075 &  0.056 \\ 
    1111 &  0.044 &  0.026 &  0.012 &  0.154 \\ 
\hline
   00000 &  0.013 &  0.011 &  0.034 &  0.028 \\ 
   00001 &  0.006 &  0.004 &  0.040 &  0.030 \\ 
   00010 &  0.009 &  0.011 &  0.053 &  0.028 \\ 
   00011 &  0.018 &  0.019 &  0.061 &  0.028 \\ 
   00100 &  0.013 &  0.011 &  0.032 &  0.053 \\ 
   00101 &  0.010 &  0.006 &  0.037 &  0.055 \\ 
   01010 &  0.019 &  0.011 &  0.042 &  0.066 \\ 
   11111 &  0.049 &  0.027 &  0.011 &  0.152 \\ 
\hline
\end{tabular}
\caption{Total variational distance between the actual distribution and the various 
approximate distribution for the case $r=2$ and $(l,k)$ 
$(3,128)$, $(4,256)$, $(5,512)$, $(6,1024)$, $(7,2048)$ and  $(8,4096)$: 
liquid theory approximation (L), Eq.~(\ref{eqn:phat}),
the liquid theory approximation normalized by an overall constant (NL), the compound
poisson approximation (CP) and the gaussian approximation (KB). }   
\end{center}
\label{dtvl3_6_r2}
\end{table}

\begin{table}
\vskip 0.25cm
\begin{center}
\begin{tabular}{||r|l|l|l|l||}
\hline
${\bf c}$  & $d^{L}_{TV}$ & $d^{NL}_{TV}$ & $d^{CP}_{TV}$ & $d^{KB}_{TV}$ \\
\hline
  000000 &  0.025 &  0.024 &  0.004 &  0.037 \\ 
  000001 &  0.003 &  0.003 &  0.028 &  0.028 \\ 
  000010 &  0.004 &  0.002 &  0.023 &  0.031 \\ 
  000011 &  0.005 &  0.006 &  0.031 &  0.031 \\ 
  000100 &  0.004 &  0.002 &  0.023 &  0.038 \\ 
  000101 &  0.004 &  0.003 &  0.029 &  0.037 \\ 
  000111 &  0.004 &  0.003 &  0.029 &  0.042 \\ 
  001001 &  0.011 &  0.012 &  0.033 &  0.041 \\ 
  010101 &  0.023 &  0.013 &  0.026 &  0.067 \\ 
  111111 &  0.052 &  0.022 &  0.015 &  0.146 \\ 
\hline
 0000000 &  0.023 &  0.022 &  0.009 &  0.040 \\ 
 0000001 &  0.022 &  0.021 &  0.006 &  0.040 \\ 
 0000010 &  0.004 &  0.002 &  0.013 &  0.031 \\ 
 0000011 &  0.018 &  0.017 &  0.004 &  0.041 \\ 
 0000100 &  0.003 &  0.002 &  0.014 &  0.034 \\ 
 0000101 &  0.010 &  0.008 &  0.007 &  0.039 \\ 
 0000111 &  0.003 &  0.002 &  0.014 &  0.037 \\ 
 0001000 &  0.003 &  0.003 &  0.017 &  0.036 \\ 
 0001001 &  0.005 &  0.003 &  0.012 &  0.040 \\ 
 0010010 &  0.005 &  0.005 &  0.017 &  0.046 \\ 
 0010011 &  0.010 &  0.007 &  0.007 &  0.055 \\ 
 0101010 &  0.025 &  0.009 &  0.012 &  0.072 \\ 
 1111111 &  0.054 &  0.020 &  0.015 &  0.140 \\ 
\hline
\end{tabular}
\caption{Total variational distance between the actual distribution and the various 
approximate distribution for the case $r=2$ and $(l,k)$ 
values of $(7,2048)$ and  $(8,4096)$: 
liquid theory approximation (L), Eq.~(\ref{eqn:phat}),
the liquid theory approximation normalized by an overall constant (NL), the compound
poisson approximation (CP) and the gaussian approximation (KB). }   
\end{center}
\label{dtvl7_8_r2}
\end{table}

Regarding the robustness of the liquid theory approximations (L) and (NL), we have
checked that going to a higher cut-off does not improve the distributions
very much. Also, it turns out that for large $\chi$ and $l$, the first order expression
for $x$ is often sufficient, however it is almost always insufficient for small 
$\chi$ and in particular when $\chi = 1$, {\it i.e.} $x$ belongs to the 
equivalence class ${\bf c} = 11\ldots1$.

Fig.~\ref{ndist_plot_k4096} shows the $n$ match distributions for $l=4$ and 
with a string length that has been increased to $k=4096$. Comparing with
the case $k=256$, Fig.~\ref{ndist_plot}, the distributions for small $\chi$ 
are more symmetric around their mean.
\vspace*{1cm}
\begin{figure}[!ht]
\includegraphics[width=16cm]{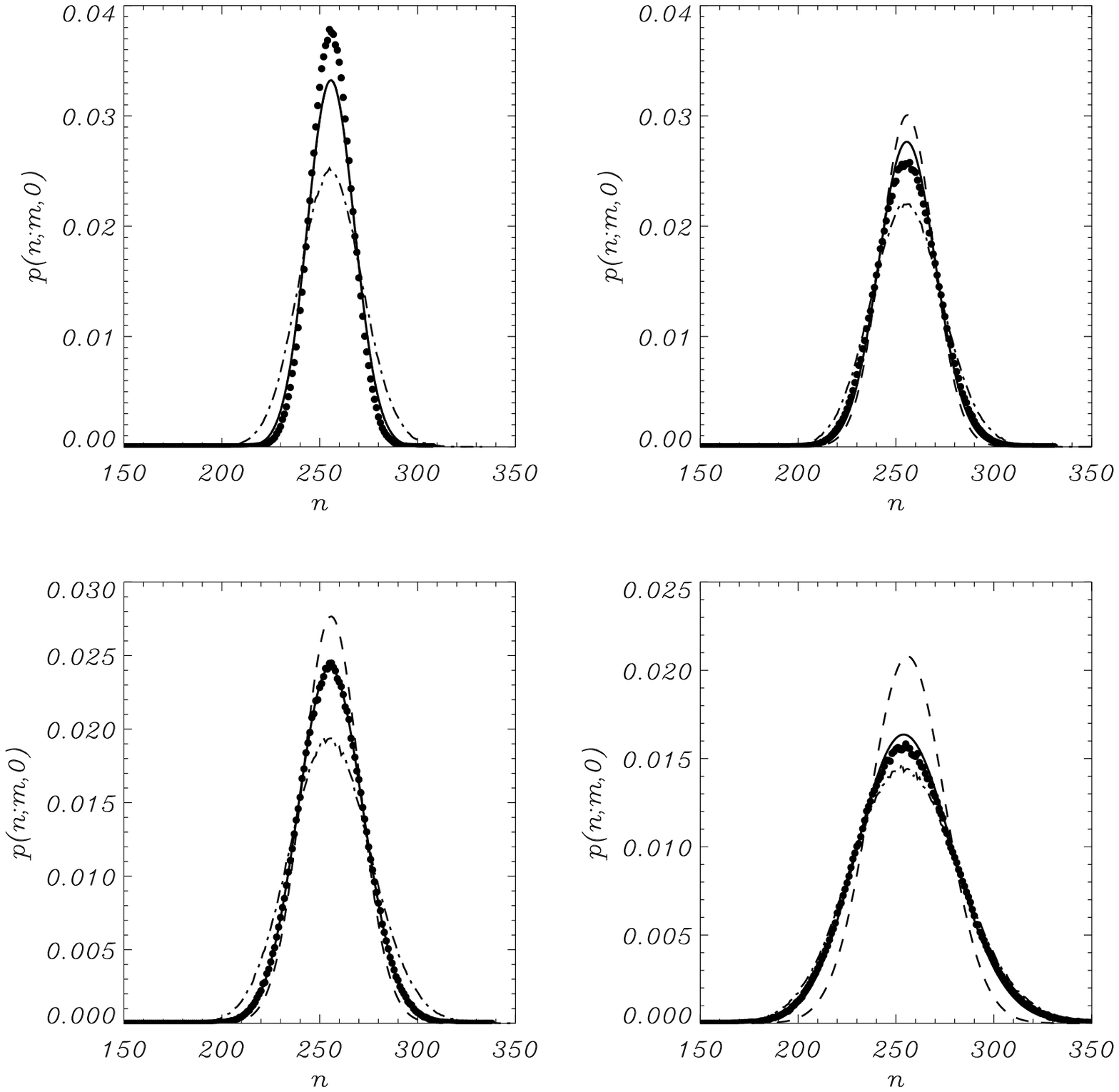}
\caption[]{The $n$-match distribution for matching a $l=4$ letter binary 
string $x$ inside a random string of length $k=4096$, for $x=0001$ (top left),
$x=1001$ (top right), $x=1010$ (bottom left) and $x=1111$ (bottom right).
The circles are the exact probabilities, the dashed and dashed-dotted lines 
correspond to the Gaussian and compound poisson approximation (see text for details). 
The solid line is the analytical result, Eq.~(\ref{eqn:phat}) normalized by an 
overall constant.
}
\label{ndist_plot_k4096}
\end{figure}
The total variatonal distances are given in the table below. Note that they are comparable 
with the values that we obtained for $k=256$, Table 3.
\begin{table}
\vskip 0.25cm
\begin{center}
\begin{tabular}{||r|l|l|l|l||}
\hline
${\bf c}$  & $d^{L}_{TV}$& $\bar{d}^{NL}_{TV}$ & $d^{CP}_{TV}$ & $d^{KB}_{TV}$ \\
\hline
     000 &  0.061 &  0.060 &  0.197 &  0.060 \\ 
     001 &  0.035 &  0.035 &  0.076 &  0.075 \\ 
     010 &  0.011 &  0.004 &  0.108 &  0.065 \\ 
     111 &  0.045 &  0.023 &  0.038 &  0.140 \\ 
\hline
\end{tabular}
\caption{Total variational distance between the actual distribution and the various 
approximate distribution for the case $r=2$, $l=4$ and $k=4096$: 
liquid theory approximation (L), Eq.~(\ref{eqn:phat}),
the liquid theory approximation normalized by an overall constant (NL), the compound
poisson approximation (CP) and the gaussian approximation (KB). }   
\end{center}
\label{dtv_k4096_l4_r2}
\end{table}

The discrepancy between actual and approximate distributions for 
${\bf c} = 000$ is persistent: it does not improve with 
increasing $\Lambda$, or going to third order
in the expansion of $x$, or by taking the stationary phase 
approximation to higher order (which turns out to be a $1/n$ expansion). 
The discrepancy for ${\bf c} = 000$ does not seem to be a finite-size effect 
either as can be seen by comparing Figs.~\ref{ndist_plot_k4096} and \ref{ndist_plot}
   
On the other hand, increasing $r$, does reduce the total variations. 
Fig.~\ref{ndist_plot_k4096r04} shows the $n$-match distribution for 
$l=4$, $m=4092$ and strings whose letters come from a $4$ letter alphabet. 
Notice that the total variation of the approximate distributions, are overall
much smaller and all three approximations yield similar results. 
In particular the deviations for ${\bf c} = 000$ have disappeared now.
Table 7 gives the corresponding variational distances:
\begin{table}
\vskip 0.25cm
\begin{center}
\begin{tabular}{||r|l|l|l|l||}
\hline
${\bf c}$  & $d^{L}_{TV}$& $\bar{d}^{NL}_{TV}$ & $d^{CP}_{TV}$ & $d^{KB}_{TV}$ \\
\hline
     000 &  0.008 &  0.008 &  0.016 &  0.030 \\ 
     001 &  0.005 &  0.003 &  0.004 &  0.034 \\ 
     010 &  0.006 &  0.007 &  0.014 &  0.036 \\ 
     111 &  0.028 &  0.013 &  0.005 &  0.080 \\ 
\hline
\end{tabular}
\caption{Total variational distance between the actual distribution and the various 
approximate distribution for the case $r=4$, $l=4$ and $k=4096$: 
liquid theory approximation (L), Eq.~(\ref{eqn:phat}),
the liquid theory approximation normalized by an overall constant (NL), the compound
poisson approximation (CP) and the gaussian approximation (KB). }   
\end{center}
\label{dtv_l4_r4}
\end{table}
Comparing with Table 3, we see indeed that for 
$r=4$ the total variational distances are overall smaller. 

It seems that for the case ${\bf c} = 000$ and $r=2$, the stationary phase approximation 
around the single point $u \approx 1$ is not capturing all the contributions to the 
probability distribution. 
\vspace*{1cm}
\begin{figure}[!ht]
\includegraphics[width=16cm]{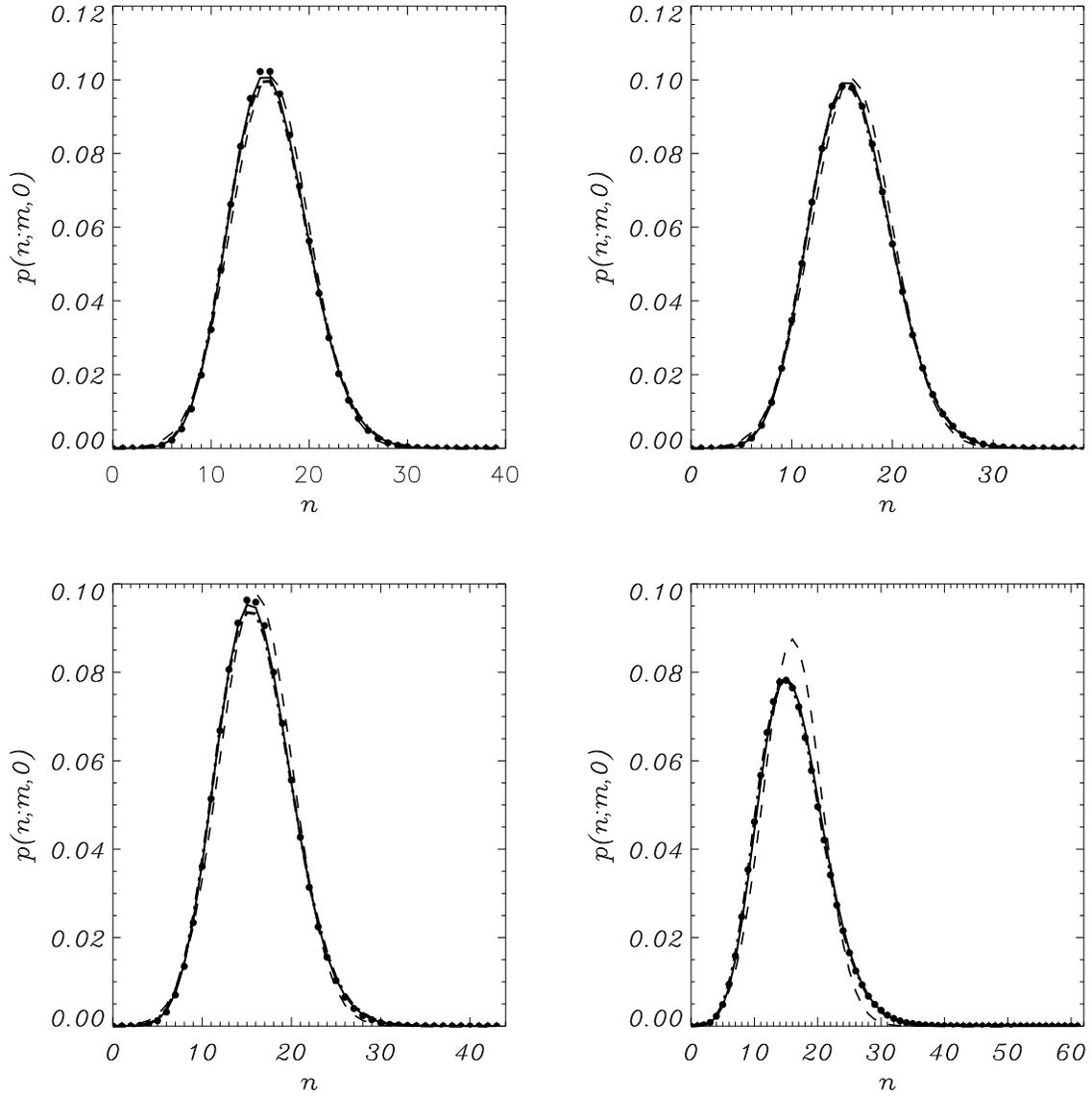}
\caption[]{The $n$-match distribution for matching a $l=4$ letter 4-ary 
string $x$ inside a random string of length $k=4096$, for $x=0001$ (top left),
$x=1001$ (top right), $x=1010$ (bottom left) and $x=1111$ (bottom right).
The circles are the exact probabilities, the dashed and dashed-dotted lines 
correspond to the Gaussian and compound poisson approximation (see text for details). 
The solid line is the analytical Eq.~(\ref{eqn:phat}).
}
\label{ndist_plot_k4096r04}
\end{figure}

Finally, we would like to remark that the expansion of $x$, Eq.~(\ref{eqn:virialx}) 
is in fact the virial expansion of the equation of state for the (discrete) lattice 
gas. The parameter $x$ is related to $z$ as $x=1/z - 1$, Eq.~(\ref{eqn:xu}). In the 
continuous 1d gas of $n$ particles in a "volume" $L$ and nearest-neighbor interactions, 
the partition function can be written as \cite{gursey,fisher}
\begin{equation}
Q(n,L) = \frac{1}{2\pi i} \oint ds e^{s L} \; \; 
D^2(s) H^{n-1}(s)
\label{eqn:partcont}
\end{equation}
where $D(s)$ and $H(s)$ are the Laplace transforms of the Boltzmann factor 
for the particle-boundary and particle-particle interactions, and Eq.~(\ref{eqn:partcont})
is the inverse Laplace transform with an appropriately chosen contour. 
For physical interactions and in the thermodynamical
limit, it turns out that the integral in the above equation 
can be evaluated by a saddle point expansion around the point $s_0$ \cite{fisher} and
as a result, it turns out that $s_0 = \beta P$, where
$\beta$ is the Boltzmann factor and $P$ is the pressure \cite{gursey}, \cite{fisher}. 
Comparing with Eq.~(\ref{eqn:pnmc_contour_app}) we see that upon discretizing 
the length of the container by letting $L = m \Delta$, 
and assuming that the interactions vary slowly with 
respect to $\Delta$, Eq.~((\ref{eqn:pnmc_contour_app}) can be recovered under the identification
\begin{equation}
e^{-s_0\Delta} = u = \frac{1}{1+x},   
\end{equation}
which for small $\Delta$ implies that x = $s_0 \Delta = \beta P \Delta$. We thus see
that the virial expansion Eq.~(\ref{eqn:virialx}) leads to a van der Waals type 
equation of state \cite{uhlenbeck}. Indeed as can be seen from Eq.~(\ref{eqn:eps1}),
$\epsilon_1$ is the {\em effective} hard-core size and the term $(n-1)\epsilon_1$ is
the total excluded "volume" due to the interaction (core + tail).   

Fig.~\ref{eq_state} shows the "$P-V$ isotherms" of the lattice gas with $l=4$, $r=2$ and 
fixed particle number $n = 15$ for the four equivalence classes ${\bf c} = 000,001,010$
and $111$ (from top to bottom). The thick solid line is the "ideal gas" law $x=n/m$.
The data points have been obtained from numerically solving Eq.~(\ref{eqn:mofx}). 
Using the approximate equation of state, Eq.~(\ref{eqn:mxeq}) give similar results
but with increasing deviations at high densities.
\vspace*{1cm}
\begin{figure}[!ht]
\includegraphics[width=16cm]{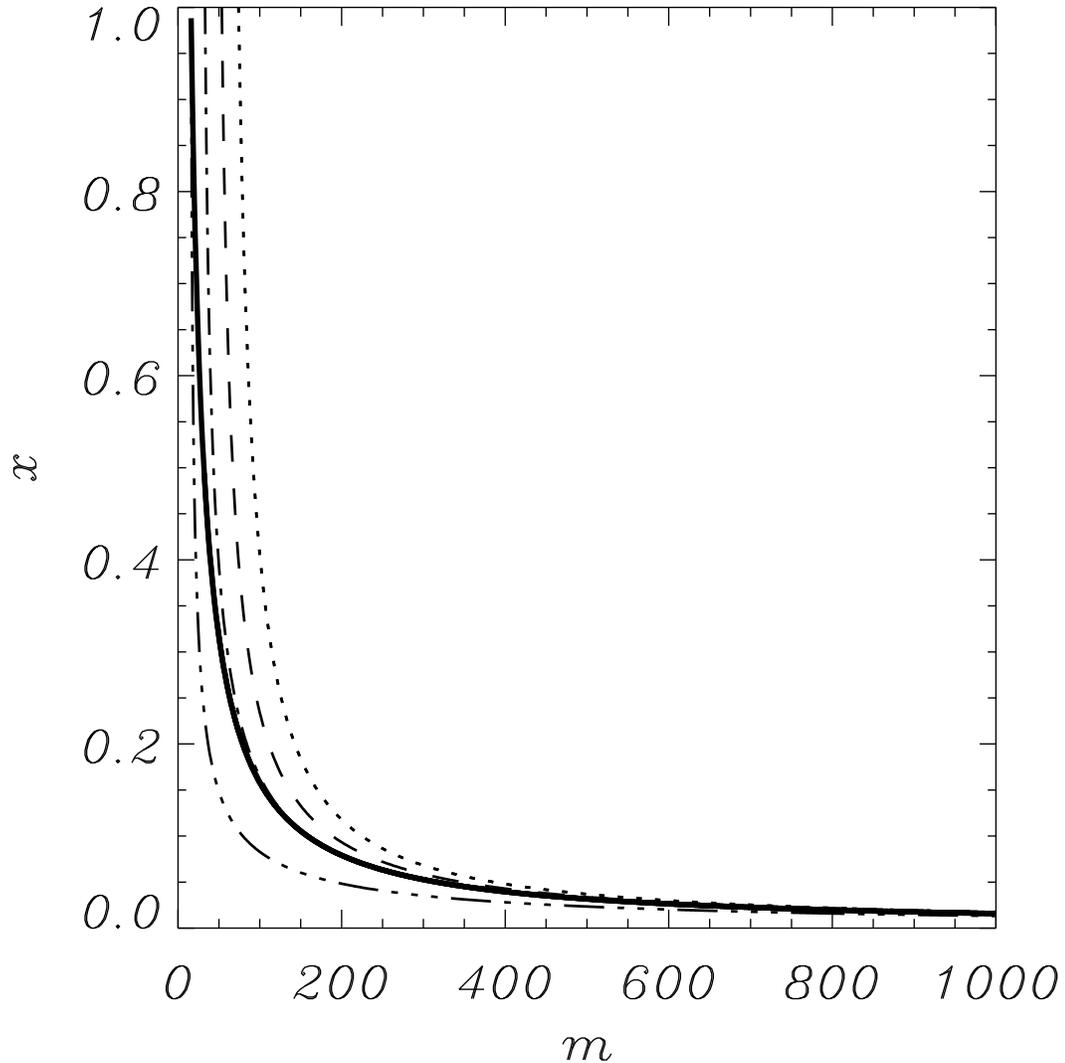}
\caption[]{The "P-V diagram" of the lattice gas with $l=4$, $r=2$ and fixed
particle number $n=15$ for the four possible interactions .
${\bf c} = 000,001,010$ and $111$ (from top to bottom). 
The thick solid line corresponds to the "ideal gas" law $x=n/m$ 
(refer to text for details).
}
\label{eq_state}
\end{figure}

\subsection{Asymptotics}

We now consider the asymptotic form of the $n$-match distributions in the limit that
the length $k=m+l$ of the random string is large. It turns out that this is 
most readily done using generating functions. We define the generating
function $p(\zeta,z;{\bf c})$ of $p(n,m;{\bf c})$  as 
\begin{equation}
p(\zeta;m,{\bf c}) = \sum_{n=0}^{\infty} p(n;m,{\bf c}) \zeta^n
\end{equation}

From Eq.~(\ref{eqn:pnmc_contour}) we thus find that
\begin{equation}
p(\zeta;m,{\bf c}) = \frac{A_1}{z_1^{m+1}} 
+  \frac{A_1}{z_1^{m+1}} \sum_{n=1}^{\infty} \left ( \zeta e^{\beta \mu} \right )^n  
\frac{1}{2\pi i} \oint_{\partial D} dz \frac{1}{z^{m+1}} \; \; D^2(z) H^{n-1}(z),
\label{eqn:pzetamc_contour}
\end{equation}
where we have used the asymptotic form $p(0;m,{\bf c}) = A_1/z_1^{m+1}$ for the $n=0$
term, since $m$ is assumed to be large. The order of summation and integration 
can be exchanged if the integrand is uniformly converging in the region of 
integration. It is not hard to show that this can be achieved for example by a 
circular path $\| z \| = R$, with a suitably chosen $R < 1$. Thus carrying out
the sum first, we obtain
\begin{equation}
p(\zeta;m,{\bf c}) = \frac{A_1}{z_1^{m+1}} 
+  \frac{A_1}{z_1^{m+1}} \zeta e^{\beta \mu}   
\frac{1}{2\pi i} \oint_{\partial D} dz \frac{1}{z^{m+1}} \; \; 
\frac{D^2(z)}{1 -  \zeta e^{\beta \mu} H(z)}.
\label{eqn:pzetamc_contour2}
\end{equation}
Substituting the approximate forms for $D(z)$ and $H(z)$, Eqs.~(\ref{eqn:dhat}) and 
(\ref{eqn:dhat}), we find
\begin{eqnarray}
&\hat{p}~(\zeta;\gamma,m,{\bf c})& = \frac{A_1}{z_1^{m+1}} \nonumber \\ 
&+&  \frac{A_1}{z_1^{m+1}}    
\frac{\zeta e^{\beta \mu}}{2\pi i} \oint_{\partial D} \frac{dz}{z^{m+1}} \; \frac{1}{1-z} \; \;
\frac{ \left [ z^\Lambda+ (1-z) \left ( D_\Lambda(z) + \Gamma(z) \right )  \right ]^2}
{ (1-z) \left [ 1 -  \zeta e^{\beta \mu} \left ( H_\Lambda(z) + \Gamma(z) \right ) \right ]  - \zeta e^{\beta \mu} z^\Lambda } . \nonumber \\ 
\label{eqn:pzetamc_contour3} 
\end{eqnarray}

Denote the expression in the denominator by $\bar{\lambda}(z;\zeta,{\bf c})$,
\begin{equation}
\bar{\lambda}(z;\zeta,{\bf c}) = (1-z) \left [ 1 -  \zeta e^{\beta \mu} \left ( H_\Lambda(z) + \Gamma(z) \right ) \right ]  - \zeta e^{\beta \mu} z^\Lambda .  
\label{eqn:lambdabar}
\end{equation}
Since $\exp(\beta \mu)$, is of order $1/r^l$ it follows that $\bar{\lambda}(z;\zeta,{\bf c})$
has a root near $z = 1$. It turns out again that this is the root closest to the origin and 
that all other roots are of order $\| z \|^\Lambda \zeta \exp(\beta \mu) \sim 1$.
Denoting the root of smallest magnitude by $\bar{z}_1$, and using the method of 
Section \ref{asySec}, a series expansion of $\bar{z}_1$ can be made. One 
finds to lowest order that 
\begin{equation}
\bar{z}_1 = 1- \frac{\zeta e^{\beta \mu} }
{1- \zeta e^{\beta \mu} H_\Lambda(1) - \zeta e^{\beta \mu} \Gamma(1)}.
\label{eqn:z1bar}
\end{equation}

The integrand in Eq.~(\ref{eqn:pzetamc_contour3}) has therefore two dominant poles 
at $z=1$ and $z=\bar{z}_1$. For large $m$, the contour integral can again be evaluated   
approximately by pushing  the countour out to infinity and keeping only the residues from 
the dominant poles (which are traversed counter-clockwise), as explained in Section \ref{asySec}. 
  
We find
\begin{equation}
\hat{p}(\zeta;\gamma,m,{\bf c}) = 
\frac{A_1}{\left (z_1 \bar{z}_1 \right )^{m+1}}
\frac{\zeta e^{\beta \mu}}{1-\bar{z}_1} 
\left ( - \frac{1}{\bar{\lambda}^\prime ((z;\zeta,{\bf c}) } \right ) 
\left [ (1-\bar{z}_1)\left (D_\Lambda(\bar{z}_1) + \Gamma(\bar{z}_1) \right ) + \bar{z}_1^\Lambda \right ]^2. 
\label{eqn:phat_largem}
\end{equation}
Notice that the $m$ dependence is entirely confined to the 
term $1/(z_1 \bar{z}_1)^{m+1}$. Thus this 
term alone is responsible for the large $m$ behavior.   
The term in the square brackets is the effect due to the 
boundaries of the string. When $m$ is 
large boundary effects should not matter and we will set 
this term to $1$. Alternatively, we
can assume that the random string is circular and 
in this case the boundary term will not arise. 

Apart from the cut-off assumption on the behavior of the tails, and the assumption of 
large $m$ leading to the $m$-asymptotic expression, Eq.~(\ref{eqn:phat_largem}), 
we have not made any assumptions on $r$ or $l$ so far. 
To proceed further, we will assume that  $1/r^l \ll 1$ so that the lowest order expressions 
for $\bar{z}_1$ and $z_1$ will provide the leading order approximation 
to Eq.~(\ref{eqn:phat_largem}).

Substituting the lowest order expression for $\bar{z}_1$, Eq.~(\ref{eqn:z1bar}) 
and noting that 
to this order 
$-\bar{\lambda}^\prime ((z;\zeta,{\bf c}) = 1 - \zeta \exp(\beta \mu) H_\Lambda(1)
- \zeta \exp(\beta \mu) \Gamma(1) $,
the result simplifies to
\begin{equation}
\hat{p}(\zeta;\gamma,m,{\bf c}) = \frac{A_1}{\left (z_1 \bar{z}_1 \right )^{m+1}}
\end{equation}
 
The compound poisson distribution arises in the limit when $m \rightarrow \infty$ and 
$\langle n \rangle $ is finite. From Eq.~(\ref{eqn:nave}) this implies that the word
length $l$ scales as $l \sim \log_r (m+1)$. From the properties of the interactions
that were derived in Section \ref{interac}, we see that the tails are very 
weak and of order $1/m$, while the core is relatively strong and of order $\log m$. 
Thus it is permissible to set $\Lambda = l$ and ignore the tails ($\Gamma = 0)$.
Note that in this limit $1/r^l \sim 1/m$ and thus to lowest order $A_1 =1 $, 
and
\begin{equation} 
e^{\beta \mu} = \frac{1}{r^l} \; \frac{1}{\left [ 1 + c(1/r) \right ]^2} 
\end{equation}

We thus obtain
\begin{equation}
\hat{p}(\zeta;\gamma,m,{\bf c}) = \left [
\left ( 1 + \frac{1}{1 + c(1/r) }\; \; \frac{1}{r^l} \right ) 
	\left ( 1-  \frac{\zeta}{r^l} \; \frac{1}{\left [ 1 + c(1/r) \right ]^2}
	\frac{1}{1-\zeta e^{\beta \mu} H_l(1) }  \right ) \right ]^{-(m+1)},
\label{eqn:almost}
\end{equation}

Further simplifications occur, noting 
that from Eq.~(\ref{eqn:hexpl0}) to order $1/r^l$ we have
\begin{equation}
\frac{1}{1 + c(1/r) } = 1 - h\left ( \frac{1}{r};{\bf c} \right ), 
\label{eqn:hcrelation}
\end{equation}
while  from Eq.~(\ref{eqn:Hdef}) we find that
\begin{equation}
e^{\beta \mu} H_l(1)  = h\left (\frac{1}{r};{\bf c} \right ).
\end{equation}
Multiplying out the product in Eq.~(\ref{eqn:almost}) and keeping only
terms to order $1/r^l \sim 1/m$, we thus obtain
\begin{equation}
\hat{p}(\zeta;\gamma,m,{\bf c}) = \left [
 1 + \frac{1}{r^l} \left ( 1- h\left (\frac{1}{r};{\bf c} \right ) \right )^2 
 \left ( \frac{1}{1 - h\left (\frac{1}{r};{\bf c} \right )} - 
-  \frac{\zeta}
{1-\zeta h \left ( \frac{1}{r};{\bf c} \right ) }  \right ) \right ]^{-(m+1)},
\end{equation}
Taking now the limit $m \rightarrow \infty$ 
such that $(m+1)/r^l = \langle n \rangle $ is finite,
the expression is readily brought to the form
\begin{equation}
\hat{p}(\zeta;\gamma,m,{\bf c}) = e^{- \sum_{j=1}^{\infty} 
\left ( 1- \zeta^j \right ) \bar{\lambda}_j } 
\label{eqn:cppoisson}
\end{equation}
with 
\begin{equation}
\bar{\lambda}_j = \langle n \rangle \left [ 1 - h\left (\frac{1}{r};{\bf c} \right )
 \right ]^2 h\left (\frac{1}{r};{\bf c} \right )^j .
\end{equation}
Eq.~(\ref{eqn:cppoisson}) is the generating function of a 
compound poisson distribution \cite{feller}
and precisely the result derived by various other methods by 
Chrysaphinou and Papastavridis \cite{chrys}, 
Geske {\it et al.} \cite{geske}, and Schbath \cite{schbath} 
in the special case of uniformly i.i.d letters. 
Also note that the CP distribution is normalized, $\hat{p}(1;\gamma,m,{\bf c}) =1 $. 

Note that setting the tails ($b \ge l$) of the interactions to zero means that given the 
next match is a distance at least $l$ away, it can occur with 
equal probability at any $b\ge l$. 
Since nearest neighbor match separations $b<l$ define an overlapping cluster, this means 
that the location of the clusters themselves, $b \ge l$, are distributed  like the arrivals 
of a poisson process \cite{chrys,reinertetal,waterman}. We therefore see that the 
liquid theory description in terms of interactions along with the separation of 
cores and tails provides an alternative and very simple explanation of this property. 
Conversely, strong tails mean that the positions of the clusters themselves are correlated 
and deviate from a poisson process (meaning that the probability of initiating a new cluster
depends on the distance from the last cluster).   
    
We now consider the limit $m \rightarrow \infty$ and 
$n \rightarrow \infty$ such that in this limit
the number density $n/(m+1) = 1/r^l$ remains constant and is small. 
In this limit the tails of the
interaction are also small, and we obtain (to lowest order in $1/r^l$)
\begin{eqnarray}
\hat{p}(\zeta;\gamma,m,{\bf c}) &=& \left [
\left ( 1 + \frac{1}{1 + c(1/r) }\; \; \frac{1}{r^l} \right ) \right ]^{-(m+1)} \nonumber \\ 
&\times& \left [ \left ( 1-  \frac{\zeta}{r^l} \; \frac{1}{\left [ 1 + c(1/r) \right ]^2}
	\frac{1}{1-\zeta e^{\beta \mu} H_l(1)  - \zeta e^{\beta \mu} \Gamma(1)}  \right ) \right ]^{-(m+1)},
\label{eqn:almost-asy}
\end{eqnarray}
Notice that if $\Lambda = \infty$, there would be nothing left for the remaining tail and thus
$\Gamma$ would be zero and we would obtain
\begin{equation}
\hat{p}(\zeta;0,m,{\bf c}) = \left [
\left ( 1 + \frac{1}{1 + c(1/r) }\; \; \frac{1}{r^l} \right ) 
	\left ( 1-  \frac{\zeta}{r^l} \; \frac{1}{\left [ 1 + c(1/r) \right ]^2}
	\frac{1}{1-\zeta  h(1/r;{\bf c})}  \right ) \right ]^{-(m+1)},
\label{eqn:almost-asy2}
\end{equation}
The normalization is given by $\hat{p}(1;0,m,{\bf c}) =1$, and using the relation   
Eq.~(\ref{eqn:hcrelation}) it is readily seen that the distribution is normalized to order
$1/r^l$. This observation immediately gives us a way to estimate $\Gamma(1)$, which must be
chosen such that the distribution is normalized to that order. We have
\begin{equation}
 e^{\beta \mu} \Gamma(1) = e^{\beta \mu} H_l(1) - h\left ( \frac{1}{r}; {\bf c} \right )  
\end{equation}
and the normalized distribution becomes
\begin{equation}
\hat{p}(\zeta;m,{\bf c}) = \left [
\left ( 1 + \left [ 1 - h\left (\frac{1}{r};{\bf c} \right ) \right ]\frac{1}{r^l} \right ) 
	\left ( 1-  \frac{\zeta}{r^l} \; 
	\frac{\left [ 1 - h\left (\frac{1}{r};{\bf c} \right ) \right ]^2} {1-\zeta  h(1/r;{\bf c})}  \right ) 
\right ]^{-(m+1)}.
\label{eqn:almost-asy3}
\end{equation}
The large $n$ limit can again be obtained using Hayman's method introduced in the previous sub-section.
Choosing $\zeta_0$ such that
\begin{equation}
n = \left. \left ( \zeta \frac{\rm d}{{\rm d}\zeta } \ln \hat{p}(\zeta;m,{\bf c}) \right ) 
\right |_{\zeta = \zeta_0}  
\end{equation}
we find to order $1 - \langle n \rangle /n$
\begin{equation}
\zeta_0 = 1 + \frac{1}{2} \; \frac{1-h(1/r;{\bf c})}{ 1+ h(1/r;{\bf c})} 
\left ( 1 - \frac{\langle n \rangle}{n} \right ),
\end{equation}
where $ \langle n \rangle$ is as defined in 
Eq.~(\ref{eqn:nave}). Using this approximation for 
$\zeta_0$, we find after a little bit of algebra that 
the distribution of $n$ around its mean is Gaussian distributed,  
\begin{equation}
\hat{p}(\zeta;m,{\bf c}) \frac{1}{\sqrt{2\pi \hat{\sigma}^2_n}} 
\exp \left (- \frac{(n - \langle n \rangle )^2}{2\hat{\sigma}^2_n} \right ), 
\end{equation}
with
\begin{equation}
\hat{\sigma}^2_n = \langle n \rangle \; \frac{1-h(1/r;{\bf c})}{ 1+ h(1/r;{\bf c})}.
\end{equation}

In concluding this section we would like to point out that our 
derivation of the CP and Gaussian asymptotic forms rests 
on determing the dominant root of $\bar{\lambda}(z;\zeta,{\bf c})$, Eq.~(\ref{eqn:lambdabar}), 
which in turn emerges as a result of introducing a cut-off $\Lambda$ and approximating the 
interactions beyond $\Lambda$. In a sense, it is the presence of the cut-off that 
simplifies the analytical treatment of the problem, since it makes explicit the separation 
of small and therefore negligible terms from the dominant ones. 
 
\section{The case of general random letter strings}
All the calculations and results presented so far, have been worked out for the case of 
uniformly and i.i.d letters of the random string. However for many applications 
this requirement is too restrictive. Letter distributions that have been considered
in the literature are non-uniform i.i.d letters and letter sequences generated by 
a Markov process. For either of the cases asymptotic results in the form of large
deviations, Gaussian and compound poisson distributions exist 
\cite{chrys,geske,fudos,regszpan,goldwater,schbath,prum,reinertetal,waterman}. 

In this section we show that the $n$-match probability associated with a broader class 
of letter distributions can be worked out using the lattice gas description introduced
in the previous section. The essential insights gained from this approach are not
changed by this generalization. The problem to be solved is still that of calculating 
the partition function of a 1d lattice gas of $n$ particles 
with nearest-neighbor interactions among themselves and the boundaries. The only 
difference is that the interactions and hence the calculations become more involved.

The required generating functions have been already derived by 
R\'egnier and Szpankowski \cite{regszpan} and we will adopt their results
to our notation. Let again ${\bf y} = (y_1,y_2,\ldots,y_k)$ be the letters of the 
random string and let ${\bf x} = (x_1,x_2,\ldots,x_l)$ be the word to be matched.  
R\'egnier and Szpankowski consider the case of i.i.d letters with arbitrary
letter distribution (Bernoulli Model) and letter sequences
generated by a one-step Markov process with transition matrix ${\bf P}$, such that
$P_{ij}$ is the transition probability $P\{y_{a+1} = i | y_a = j\}$, 
${\bf \pi} = (\pi_1,\pi_2,\ldots,\pi_r)$ is the stationary letter distribution satisfying
${\bf \pi} {\bf P} = {\bf \pi}$, 
and the stationary matrix $\Pi$ is the matrix whose $r$ rows
are ${\bf \pi}$ (Markov Model). 

Given any subsequence of letters $y_{a+1}, y_{a+2},\ldots, y_{a+l}$, 
denote by $p({\bf y}_{a,l})$  the probability of encountering 
${\bf y}_{a,l}$, without any conditions on the letters preceeding or following it.
Likewise, denote by $p({\bf x})$ the probability of generating 
the word ${\bf x}$. The generating function of the $n$-match probability 
is given by \cite{regszpan}: 
\begin{equation}
p(n;z,{\bf c}) =  p({\bf x}) \tilde{d}^2(z;{\bf c}) \tilde{h}^{n-1}(z;{\bf c}),
\label{eqn:pnztilde}
\end{equation}
with
\begin{equation}
\tilde{d}(z;{\bf c}) = \frac{1 - \tilde{h}(z;{\bf c})}{1-z} 
= \frac{1}{p({\bf x})} \; \frac{1}{\lambda(z;{\bf c})},
\label{eqn:dztilde}
\end{equation}
\begin{equation}
\tilde{h}(z;{\bf c}) = 1 - \frac{1}{p({\bf x})} \; \frac{1 - z}{\lambda(z;{\bf c})},
\label{eqn:hztilde}
\end{equation}
and 
\begin{equation}
\lambda(z;{\bf c}) = z^l + \frac{1}{p({\bf x})} (1-z) \left [ 1 + \tilde{c}(z) 
+ \frac{p({\bf x})}{\pi(x_1)} T(z) z^l \right ],
\label{eqn:lz}
\end{equation}
In the last equation $\pi(x_1)$ is the steady state 
probability of encountering the letter $x_1$ and $T(z)$ is the generating 
function for the steady-state transition probability from the end of one 
word match to the beginning of the next word match as a function of the gap
length between the two words (for the Bernoulli Model $T(z) = 0$). 
The generating function 
$\tilde{c}(z)$ is defined as
\begin{equation}
\tilde{c}(z) = \sum_{b=1}^{l-1} c_b p({\bf x}_{1,b}) z^b,
\label{eqn:cztilde}
\end{equation}
where $c_b({\bf x})$ are the bit-vectors associated with the word ${\bf x}$.
Note that an overall factor of $z^l$ in the definition of $p(n;z,{\bf c})$ 
in \cite{regszpan} is absent, since the generating function 
$p(n;z,{\bf c})$, as defined above, 
corresponds in our case to $p(n;m,{\bf c})$, where $m = k-l$ is the effective 
length of the string.   
 
Comparing with the corresponding equations of the uniformly distributed 
random letter case, Eqs.~(\ref{eqn:pnzdef}), (\ref{eqn:hzexplicit}), 
(\ref{eqn:dzexplicit}) and (\ref{eqn:lambdazdef}), we see that the form 
of the equations as well as the relationships between the generating 
functions are identical.

In particular, all recursions can be recovered by making the replacements
$h_a/r^a \rightarrow \tilde{h}_a$, $d_a/r^a \rightarrow \tilde{d}_a$, 
and $c_a/r^a \rightarrow \tilde{c}_a$ so that 
$h(z/r) \rightarrow \tilde{h}(z)$ etc.    

The Markov property introduces the additional complication that 
one has to propagate the end of one word match at $a_i$ to the 
beginning of the next match at $a_{i+1}$ through the $(a_{i+1} - a_i -l)$-step 
steady-state transition probability. 

R\'egnier and Szpankowski have also proven that the polynomial 
$\lambda(z;{\bf c})$ has at least one real root and that all roots 
have $ \| z \| \ge 1$, as in the 
case of uniform letter distributions.
The asymptotic behavior of $\tilde{h}$ and $\tilde{d}$ is again due to 
the root closest to $z=1$. 
  
As can be seen from Eq.~(\ref{eqn:lz}), for $p({\bf x})$ small, the root closest
to $z=1$ is located at roughly
\begin{equation}
z_1 \approx 1 + \frac{p({\bf x})}{ 1 + \tilde{c}(1) 
+ \frac{p({\bf x})}{\pi(x_1)} T(1) } 
\end{equation}
and all other roots are roughly located at $\| z \|  \sim 1/p({\bf x})$.  
Recall that in the case of uniformly distributed letters, 
$p({\bf x}) = 1/r^l \le 1/2$. For the general letter distributions, both the 
distribution as well as the word ${\bf x}$ can be chosen arbitrarily and 
thus there is no constraint on the values that $0 \le p({\bf x}) \le 1$ can take.  
This means in particular that there is a broader class of possible interactions. 

Defining again the effective particle-particle interaction as 
\begin{equation}
e^{-\beta U(b)} = \frac{\tilde{h}(b)}{\tilde{h}_{asy}(b)},
\end{equation}
${\tilde{h}_{asy}(b)}$ is readily worked out as
\begin{equation}
\tilde{h}_{asy}(b) = p({\bf x}) \frac{A_1}{z_1} \; 
\left [ \frac{(z_1 - 1)^2}{p({\bf x})} \right ] \; \left ( \frac{1}{z_1} \right )^b 
\equiv e^{\beta \mu} \left ( \frac{1}{z_1} \right )^b 
\end{equation}
where $A_1(z_1 - 1) = -1/\lambda^\prime(z_1)$, {\it cf.} Eq.~(\ref{eqn:A1def}).
We thus obtain for the particle-particle interaction
\begin{equation}
\beta U(b) = - \ln \tilde{h}(b) - b \ln z_1 + \ln p({\bf x}) + \beta U_0,
\label{eqn:Uppcore_gen}
\end{equation}
where 
\begin{equation}
\beta U_0 = \ln \left [ \frac{A_1}{z_1} \; \left ( \frac{z_1 - 1}{p({\bf x})} \right )^2 \right ] . 
\end{equation}
If $p({\bf x}) \ll 1$, $\beta U_0$ is a constant of order 
$p({\bf x})$, since the argument of the logarithm
is of order 1 to the same order.

In the core-region $b<l$, the non-zero values of $\tilde{h}$ are still determined by the bit-vector
${\bf c}$ associated with ${\bf x}$ and we find analogous to Eq.~(\ref{eqn:hcore}) that
\begin{equation}
\tilde{h}(b) =  \left \{ \begin{array}{ll} 
            c_b p({\bf x}_{1,b}), & \mbox {if $\chi$ does not divide $b$,} \\ 
            p({\bf x}_{1,\chi}), & \mbox{if $b = \chi $,} \\
            0, & \mbox{otherwise,} \\
\end{array} \right.
\label{eqn:hcore_gen} 
\end{equation}
where $\chi$ is the fundamental period associated with ${\bf c}$ that was 
defined at the end of Section \ref{bit-sect} and by definition,
$\tilde{h}(0) = 0$. 

We see that the interaction is $+\infty$, whenever $c_b = 0$. This is
certainly the case for $b < \chi$. The interaction in the core-region is given 
by 
\begin{equation}
\beta U(b) = - \ln c_b - b \ln \left [ z_1 p^{1/b}({\bf x}_{1,b} ) \right ] + \ln p({\bf x}) + \beta U_0,
\label{eqn:Uppcore_gen2}
\end{equation}
Comparing the above with the uniform letter distribution case, Eq.~(\ref{eqn:Uppcore}), 
we see similarities as well as differences: the argument of the logarithm in square 
brackets is no longer necessarily smaller than one, but can depend on the subtle 
interplay of the overall word matching probability $p({\bf x})$ (which determines $z_1$) 
with the (smaller) probabilities of matching the subwords $p({\bf x}_{1,b})$. 
Thus such cases 
require more care. Assuming that $p({\bf x})$ is sufficiently small so that the expression 
in square brackets is smaller than one, it is again the case that the energy of the core
region is of order $\ln p({\bf x})$ and increases with $b$. Under the same assumptions, the 
characteristic energy of the tail region can be worked out and one finds that it goes like
$p({\bf x})$ similarly to the line of reasoning in Section \ref{interac}. 

Thus we see that by suitably choosing the set of probabilities $ p({\bf x}_{1,b})$, for 
$b=1,2\ldots,l$ the strengths of the core and tail of the interactions can be varied and
can possibly move the distribution functions into a regime where approximations 
ignoring the contributions from the tails (such as the compound poisson approximation) 
are inappropriate. 
    
Also note that by letting $p({\bf x})$ to be arbitrarily close to one, the difference of  
magnitudes between the root closest to the origin $z_1$ and the other roots can be made to 
vanish. Since the attenuation length of the tail of the interactions depends on the 
separation of these roots, we see that in this limit $z_1$ ceases to be the dominant 
root. For the interactions this means an increasingly more slowly decaying tail as the 
two roots approach each other. In such a regime the tails of the interactions 
should become very important and thus cannot be neglected. 

It is possible that for certain distributions and choices of words, this can 
cause a break-down of the liquid theory approach, which essentially 
is a perturbation theory and it would be interesting
to find out if and how this can happen. Strong tails will certainly affect the quality
of approximations such as the compound poisson distribution, which was based on the  
assumption that tails can be ignored, which turned out to be equivalent to 
assuming a poissonian distribution of cluster locations, 
as explained in Section \ref{asySec}.
 
We will further discuss these points in the Discussion section below.
 
\section{Discussion}

We have presented a new approach to calculating the probability distribution for 
the number of matches of a given word inside a random string of letters. 
Our approach rests on the observation that the exact expression for such a distribution
can be interpreted as the partition function of an $n$-particle system on a 
linear lattice, with pairwise nearest neighbor interactions. By exploiting this
analogy and focusing on the generic properties of the interaction, we have been 
able to set up a virial expansion for the equation of state of this lattice 
gas and thereby obtained an analytical expression for the $n$-match probability
distribution, which besides extrapolating between the known asymptotic forms, 
also provides a good approximation in the intermediate regimes.

The identification and subsequent analysis of the effective interactions in the 
lattice gas description turns out to be key in our solution of this problem. 
The interactions are characterized by a strong core-region of the size of the 
word-length followed by a relatively weak and exponentially decaying tail. 
Although we have carried out the detailed analysis for the special case of 
uniform letter distributions, we showed in Section IV, that our 
method is readily extended to the broader class of distributions, such as 
non-uniform letter distributions and random letter sequences generated by
a Markov process. Regardless of the underlying stochastic process for the
random string, the generic feature of the interactions are still the same, 
namely a relatively strong core and a weaker tail and our approach should be 
readily applicable to these types of problems as well.   

We should also point out that our method of approach bears some 
similarity with the work of R\'egnier and Szpankowski \cite{regszpan}, who
also use generating functions in their approach to this problem. Our 
approach is however distinct in at least one crucial point: In the cited 
work, upon deriving the generating functions
for the $n$-match distribution, Eqs.~(\ref{eqn:pnztilde}), (\ref{eqn:dztilde}),
(\ref{eqn:hztilde}) and (\ref{eqn:lz}), the authors perform a 
Laurent expansion of the generating function around its dominant $n+1$ order 
pole at $z_1$. Such an expansion is asymptotic in the interactions 
and runs the risk of capturing more accurately the tail of the 
interaction rather than its core (or at least many terms must be kept in order
to capture the core part to a sufficient degree of accuracy \cite{wilf}). 
The approximation scheme presented here precisely avoids this 
by introducing a cut-off distance $\Lambda$ and keeping 
the {\em exact} interaction upto $\Lambda$, while approximating 
the interactions only beyond $\Lambda$. As we have shown, this is
easily done, since the structure of the core of the interaction ($b < l$) 
directly follows from the overlap properties of the string to be matched. 
Our analysis also shows that since the core part of the interaction 
is typically stronger than the exponentially decaying tail, keeping the core  
is crucial in determining the global properties of the distribution. 
Moreover, our approach allows us to understand approximations such 
as the compound distribution as being applicable in a regime where 
the tails of the interaction can be neglected and only the core is kept.
This also highlights the relative importance of the core part of the 
interaction with respect to its tail.
 
Lastly, we would like to remark that our treatment of interactions,
by separating out its strong and short-ranged core from its 
weak tail, is actually not new. Interactions with a strong core
and an exponentially decaying tail are known as Kac potentials,
named after M. Kac, who along with co-workers studied one-dimensional 
particle systems with such interactions (continuum and lattice 
version) in considerable detail,
as part of an effort to understand the liquid-gas 
transition in the context of the van der Waals equation of
state \cite{kac,kacetal,hemleb} (for an overview, see
the review article by Hemmer and Lebowitz \cite{hemleb}).

Such systems are interesting, since they lead to phase transitions
in the limit when the characteristic decay length of the interaction
tends to infinity \cite{kac,kacetal,hemleb} (for an overview of phase transitions in one 
dimensions see, the review article by Griffiths \cite{griffiths}). 
The similarity of such systems with the 
string matching problem is at hand, since one can make the 
interactions to decay as slowly as one wishes by 
choosing a suitable random letter distribution and string ${\bf x}$ to 
be matched such that the dominant poles of the generating function 
of the $n$-match distribution function become arbitrarily 
close to each other. It would therefore be of interest 
to see whether the distribution functions in this regime
can be calculated using the more sophisticated techniques,  
such as integral equations and operator methods, 
which have been introduced particularly for the purpose 
of dealing with such types of interactions \cite{hemleb}.

\vskip 1cm
{\bf Acknowledgments}
I would like to thank Ay\c{s}e Erzan for both initially bringing to my attention 
the string matching problem and later pointing out the connection with Kac 
potentials.  This work was supported in part by the Nahide and Mustafa 
Saydan Foundation and T\"ubitak, the Turkish Science and Technology 
Research Council.
\vskip 2cm
 
\appendix

\end{document}